\documentclass[%
aps,pre,amsmath,amssymb,showpacs,reprint,longbibliography,
tightenlines
]{revtex4-2}

\usepackage{mathtools}
\usepackage{color}
\usepackage{bigints}
\usepackage{multirow}
\usepackage{graphicx}
\usepackage{dcolumn}
\usepackage{bm}
\usepackage{subfigure}
\usepackage{hyperref}
\usepackage{multirow}
\usepackage{mathtools}

\newif\iffigure
\figurefalse
\figuretrue
\allowdisplaybreaks[3]
\usepackage{cprotect}

\def\kakuchoushi{.pdf}
\def\SI{Supporting Infomation}
\def\num{\mathrm{num}}
\def\ana{\mathrm{ana}}
\def\spi{\sqrt{\pi}}
\def\erfc{\mathrm{erfc}}
\def\ep{\varepsilon}
\def\dd{\mathrm{d}}
\def\T{\mathrm{T}}

\def\rnd#1#2{\frac{\partial #1}{\partial #2}}
\def\drnd#1#2{\dfrac{\partial #1}{\partial #2}}
\def\a{\mathrm{a}}
\def\ave{\mathrm{ave}}
\def\dif{\mathrm{dif}}
\def\b{\mathrm{b}}
\def\c{\mathrm{c}}
\def\u{\mathbf{u}}
\def\F{\mathbf{F}}
\def\f{\mathbf{f}}
\def\v{\mathbf{v}}

\def\x{\mathrm{x}}
\def\y{\mathrm{y}}
\def\m{\mathrm{m}}

\def\zR{z_{\mathrm{R}}}
\def\tzR{\tilde{z}_{\mathrm{R}}}
\def\tD{\tilde{D}}
\def\tnabla{\tilde{\nabla}}
\def\um{\textmu m}
\def\red#1{\textcolor{black}{#1}}
\def\zf{z_{\mathrm{f}}}
\def\Hf{H_{\mathrm{f}}}
\def\kf{\kappa_{\mathrm{f}}}
\def\Af{A_{\mathrm{f}}}
\def\Rey{\mathit{Re}}

\def\mI{\mathcal{I}}
\def\mIx{\mathcal{I}^{(\x)}}
\def\bmIx{\bar{\mathcal{I}}^{(\x)}}
\def\mIa{\mathcal{I}^{(\a)}}
\def\mIb#1{\mathcal{I}^{(\b #1)}}
\def\bmIa{\bar{\mathcal{I}}^{(\a)}}
\def\bmIb#1{\bar{\mathcal{I}}^{(\b #1)}}
\def\mIs{\mathcal{I}_s}
\def\mJ{\mathcal{J}}
\def\mJxy{\mJ^{(\x,\y)}}
\def\bmJxy{\bar{\mJ}^{(\x,\y)}}

\def\mJxc{\mJ^{(\x,\c)}}

\def\mJxd#1{\mJ^{(\x,\dd #1)}}

\def\bmJxc{\bar{\mJ}^{(\x,\c)}}

\def\bmJxd#1{\bar{\mJ}^{(\x,\dd #1)}}

\def\mA{\mathcal{A}}
\def\mB{\mathcal{B}}
\def\wm#1{w_{\m #1\relax}}
\def\twm#1{\tw_{\m #1\relax}}

\def\tIm#1{\tI_{\m #1\relax}}

\def\dm#1{\delta_{\m #1\relax}}
\def\dTa{\delta_{\T}^{(\a)}}
\def\dTx{\delta_{\T}^{(\x)}}
\def\dTb#1{\delta_{\T}^{(\b #1)}}
\def\dTbj{\delta_{\T }^{(\b j)}}
\def\duc{\delta_{\mathrm{u}}^{(\c)}}
\def\dudj{\delta_{\mathrm{u}}^{(\dd j)}}
\def\dud#1{\delta_{\mathrm{u}}^{(\dd #1)}}
\def\duy{\delta_{\mathrm{u}}^{(\y)}}
\def\tm#1{\tau_{\m #1\relax}}
\def\t#1{\tau_{#1\relax}}
\def\ta#1{\tau_{#1}^{(\a)}}
\def\tb#1#2{\tau_{#1}^{(\b #2)}}
\def\tx#1{\tau_{#1}^{(\x)}}
\def\btx#1{\bar{\tau}_{#1}^{(\x)}}
\def\btxj{\bar{\tau}_{j}^{(\x)}}

\def\Z{\frac{\tz}{\tH}}

\def\SSxy{\mathsf{S}^{(\x,\y)}}
\def\SSac{\mathsf{S}^{(\a,\c)}}
\def\SSbc#1{\mathsf{S}^{(\b #1,\c)}}
\def\SSxd#1{\mathsf{S}^{(\x,\dd #1)}}

\def\VVxy{\mathsf{V}^{(\x,\y)}}
\def\VVac{\mathsf{V}^{(\a,\c)}}
\def\VVbc#1{\mathsf{V}^{(\b #1,\c)}}
\def\VVxd#1{\mathsf{V}^{(\x,\dd #1)}}
\def\Cxy{C^{(\x,\y)}}
\def\Cxc{C^{(\x,\c)}}
\def\Cac{C^{(\a,\c)}}

\def\Cxdj{C^{(\x,\dd j)}}

\def\Bx#1{\mathsf{B}_{#1}^{(\x)}}

\def\EEx{\mathsf{E}^{(\x)}}
\def\EEa{\mathsf{E}^{(\a)}}
\def\EEb#1{\mathsf{E}^{(\b #1)}}

\def\GGx{\mathsf{G}^{(\x)}}
\def\GGa{\mathsf{G}^{(\a)}}
\def\GGb#1{\mathsf{G}^{(\b #1)}}
\def\HH{\mathsf{H}}
\def\QQ{\mathsf{Q}}
\def\PP{\mathsf{P}}
\def\up{\uparrow}
\def\lw{\downarrow}
\def\asfH{\mathsf{a}_H}
\def\asf#1{\mathsf{a}_{#1}}
\def\bsfH{\mathsf{b}_H}
\def\bsf#1{\mathsf{b}_{#1}}
\def\dIH{\mathsf{I}_H^\prime}
\def\IH{\mathsf{I}_{H}}
\def\Iz{\mathsf{I}_{0}}
\def\dIz{\mathsf{I}_{0}^\prime}
\def\dJH{\mathsf{J}_H^\prime}
\def\JH{\mathsf{J}_{H}}
\def\Jz{\mathsf{J}_{0}}
\def\dJz{\mathsf{J}_{0}^\prime}
\def\uac{\u^{(\a,\c)}}
\def\uad#1{\u^{(\a,\dd #1)}}
\def\ubc#1{\u^{(\b #1,\c)}}
\def\ubd#1#2{\u^{(\b #1,\dd #2)}}
\def\uxy{\u^{(\x,\y)}}
\def\busxy{\bu_s^{(\x,\y)}}

\def\busxc{\bu_s^{(\x,\c)}}
\def\busxd#1{\bu_s^{(\x,\dd #1)}}
\def\urxy{u_r^{(\x,\y)}}
\def\burxy{\bu_r^{(\x,\y)}}
\def\uzxy{u_z^{(\x,\y)}}
\def\buzxy{\bu_z^{(\x,\y)}}
\def\tpac{\tp^{(\a,\c)}}
\def\tpad#1{\tp^{(\a,\dd #1)}}
\def\tpbc#1{\tp^{(\b #1,\c)}}
\def\tpbd#1#2{\tp^{(\b #1,\dd #2)}}
\def\tpxy{\tp^{(\x ,\y)}}
\def\btpxy{\bar{\tp}^{(\x ,\y)}}
\def\Hm#1{H_{\m #1\relax}}
\def\Hmj{H_{\m j}}
\def\Tm#1{T_{\m #1\relax}}
\def\Am#1{A_{\m #1\relax}}
\def\Aa{A^{(\a)}}
\def\Ab#1{A^{(\b #1)}}
\def\Abj{A^{(\b j)}}
\def\km#1{\kappa_{\m #1\relax}}
\def\kmj{\kappa_{\m j}}
\def\k#1{\kappa_{#1\relax}}
\def\kj{\kappa_{j}}
\def\tkj{\tilde{\kappa}_{j}}
\def\tk#1{\tilde{\kappa}_{#1}}
\def\tkmj{\tilde{\kappa}_{\m j}}
\def\tkm#1{\tilde{\kappa}_{\m#1}}

\def\s#1{^{(#1)}}

\def\sl{^{(\ell)}}

\def\slm#1{^{(\ell-#1)}}
\def\hf{\frac{1}{2}}
\def\dhf{\displaystyle\frac{1}{2}}
\def\subp{|_{\tz=\hf}} 
\def\subm{|_{\tz=-\hf}} 
\def\bsubp{\Big|_{\tz=\hf}} 
\def\bsubm{\Big|_{\tz=-\hf}} 
\def\subpd{|_{z=0}} 
\def\submd{|_{z=-\Hm1}} 
\def\bsubpd{\Big|_{z=0}} 
\def\bsubmd{\Big|_{z=-\Hm1}} 
\def\Dp{\Delta_{\parallel}}
\def\tDp{\tilde{\Delta}_{\parallel}}
\def\betac{\beta^{(\c)}}
\def\Kdj{K^{(\dd j)}}
\def\Kd#1{K^{(\dd #1)}}

\def\tp{\tilde{p}}
\def\tr{\tilde{r}}
\def\tz{\tilde{z}}
\def\tzs{\tilde{z}_s}

\def\tw{\tilde{w}}

\def\tI{\tilde{I}}

\def\tH{\tilde{H}}
\def\tHj{\tilde{H}_j}
\def\tHmj{\tilde{H}_{\m j}}
\def\tHm#1{\tilde{H}_{\m #1}}
\def\tD{\tilde{D}}

\def\bt{\bar{\tau}}
\def\bu{\bar{u}}

\def\barf{\bar{f}}

\newcommand{\tausx}{\bt_s^{(\x)}}
\newcommand{\tausa}{\bt_s^{(\a)}}
\def\tausb#1{\bt_s^{(\b #1)}}

\begin{document}

\preprint{\today, ver.~\number\time}

\newcommand{\titleA}{
}

\newcommand{\titleB}{
Semi-analytical model of optothermal fluidics in a confinement
}

\title{
\titleB
}

\author{Tetsuro Tsuji}
\email{tsuji.tetsuro.7x@kyoto-u.ac.jp; corresponding author}
\author{Shun Saito}%
\author{Satoshi Taguchi}%
\affiliation{%
Graduate School of Informatics, Kyoto University, Kyoto 606-8501, Japan
}%

\date{\today}

\begin{abstract}
\noindent 
In this paper, we provide the semi-analytical solution of the temperature and flow fields of a fluid confined in a narrow space between two parallel plates. The temperature increase is triggered by photothermal effects of fluids and/or boundaries due to the absorption of a focused Gaussian beam irradiated perpendicular to the fluid film, and then the temperature variation induces the flow fields through a buoyancy force and/or thermo-osmotic slip. The semi-analytical solution to this optothermal fluidic system is validated by comparing with the results of numerical simulation, and is applied to typical optothermal fluidic problems. 
In particular, the optothermal trap of nanoparticles observed in our previous experiment [T. Tsuji, et al., Electrophoresis, 42, 2401 (2021)] is investigated in terms of thermophoretic force and flow drag that are obtained semi-analytically. 
The semi-analytical solution can be shared through open-source codes that are available to researchers without the background of fluid mechanics. 
\end{abstract}

\maketitle


\section{\label{sec:intro}Introduction}

\begin{figure}[bt]
    \centering
    \includegraphics[width=0.7\linewidth]{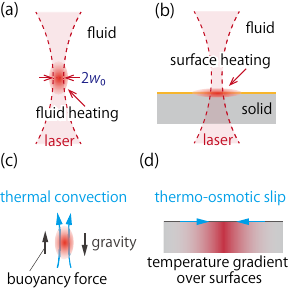}
    \caption{Photothermal effects of (a) fluids and (b) surfaces. (a) Three-dimensional heat source is localized at a laser focus where a beam intensity is high. (b) Two-dimensional heat source is localized in a thin-film attached to the surface of a solid (i.e., a fluid channel wall). Schematic of thermally-induced flows: (c) thermal convection and (d) thermo-osmotic slip.}
    \label{fig:heating}
\end{figure}

Photothermal effect is an energy conversion from light to heat through light absorbing to materials, and has two remarkable features from the view point of microfluidics and nanofluidics.  
First, since the efficiency of the conversion is dependent on the combination of materials and the wavelength of light, selective heating in a fluid system can be realized by a suitable choice of the light source, a working fluid, and a channel design. 
Second, light can be easily localized in microscale by focusing a laser, or even smaller nanoscale beyond the diffraction limit by utilizing near-field optics and thermoplasmonics \cite{Baffou2017}. 
These two features enable non-contact local heating in a microfluidic system, leading to fine temperature shaping and resulting flow generation toward applications such as pumping \cite{Ciraulo2021} and molecular manipulation \cite{Braun2002}.
The heating methods can be roughly divided into three: (i) bulk heating (or fluid heating) \cite{Braun2002,Duhr2005,Duhr2006b,Weinert2009,Cordero2009,Liu2010,Maeda2012,Riviere2016,Tsuji2018a,Tsuji2021,Ruzzi2023,Mamuti2021,Mamuti2024,Kumari2012,Motosuke2012}, (ii) boundary heating (or surface heating) 
\cite{Braun2013,Chen2015c,Kotnala2019a,Fraenzl2022,Wang2022b,Zhou2023,Kollipara2023,Chen2024,Meng2024,Chen2024a,Jiang2024,Jiang2009,Yu2015,Simon2023a,FloresFlores2015,ZentenoHernandez2020,Bregulla2016,Setoura2017,Hayashi2021,Donner2011,Shoji2014,Namura2015,Yang2023,Lu2021}, and (iii) dispersion heating \cite{Iida2016,Jin2018,Shi2023,Chen2016a,Bruot2019,Paul2022a,Schmidt2018,Fraenzl2021,Zhong2019,Qian2020,Amaya2024}. The cases of (i) and (ii) are both considered in this paper and thus schematically shown in Figs.~\ref{fig:heating}(a) and \ref{fig:heating}(b), respectively. 

Despite the increasing number of researches using photothermal effects recently, a systematic analytical method to investigate the temperature and flow fields is absent; the goal of this paper is to give an easy-access and instant computational tools that are fully open to researchers with/without the background of fluid mechanics.

Before going into the description of the framework of the present paper, let us first overview the methods of heating using photothermal effects. Fluid heating [Fig.~\ref{fig:heating}(a)] is enabled by a wavelength $\lambda$ of a near-infrared light $\approx 1.5$~\textmu m \cite{Braun2002,Duhr2005,Duhr2006b,Weinert2009,Cordero2009,Liu2010,Maeda2012,Riviere2016,Tsuji2018a,Tsuji2021,Ruzzi2023} for a water solution. The absorption coefficient at $\lambda\approx 1.5$~\textmu m is about $2\times 10^3$ m$^{-1}$ and is much higher than, e.g., $4\times10^{-2}$ m$^{-1}$ of visible wavelength $\lambda=532$~nm. 
The absorption at $\lambda\approx 2.0$~\textmu m \cite{Mamuti2021} or 3.0~\textmu m \cite{Mamuti2024} is more significant although experiments will be more delicate due to limited choices of optical components. One can add some light-absorbing molecules to solvent \cite{Kumari2012,Motosuke2012} when a light source with suitable wavelength is not available. 
The simplest case of surface heating [Fig.~\ref{fig:heating}(b)] uses a thin metal film of several tens of nanometer thickness as a light absorber; Au films \cite{Braun2013,Chen2015c,Kotnala2019a,Fraenzl2022,Wang2022b,Zhou2023,Kollipara2023,Chen2024,Meng2024,Chen2024a,Jiang2024}, Cr films \cite{Jiang2009,Yu2015,Simon2023a}, and amorphous silicon layers \cite{FloresFlores2015,ZentenoHernandez2020} were deposited on fluid channel boundaries (e.g., a glass substrate). Alternatively, Au particle \cite{Bregulla2016,Setoura2017} or Janus particle \cite{Hayashi2021} fixed on boundaries, plasmonic \cite{Donner2011,Shoji2014,Namura2015} or dielectric \cite{Yang2023} nanostructures fabricated on substrates, graphene oxide-doped polydimethylsiloxane (PDMS) \cite{Lu2021} are also available as light-absorbing boundaries. 
Note that most cases of boundary heating are surface heating, i.e., the light absorption occurs at the interface of the solid and the fluid [Fig.~\ref{fig:heating}(b)]. 
Alternatively, light-absorbing dispersed particles, such as Au nanoparticles \cite{Iida2016,Jin2018,Shi2023}, Au nanorods \cite{Chen2016a}, titania metal-oxide particles \cite{Bruot2019}, polystyrene (or silica) particles containing iron oxide \cite{Paul2022a} (or \cite{Schmidt2018}), melamine formaldehyde particles covered by the uniformly-distributed Au nanoparticle \cite{Fraenzl2021}, core/shell magnetic polystyrene particle \cite{Zhong2019}, single walled carbon nanotube \cite{Qian2020} are also available. 
In the case of the heating of dispersed particles, a generated heat may bring the particles to the heated spot, resulting in additional heating and the positive feedback loop to accumulate the particles \cite{Chen2016a,Jin2018,Amaya2024}. 
The choice of heating methods strongly depends on the wavelength of available light sources, nanofabrication resources, and required power efficiency. 
Recent review papers \cite{Chen2019,Chen2021,Kollipara2023a} are also useful to overview the variety of heating methods, where main applications are the manipulation of nanomaterials, exploring so-called optothermal manipulation.

Generated heat inherently induces fluid flows such as thermal convection [Fig.~\ref{fig:heating}(c)] and thermo-osmotic slip [Fig.~\ref{fig:heating}(d)]. Thermal convection (or natural convection) is a buoyancy-driven flow due to the thermal expansion of a fluid and is usually modeled by using the Boussinesq approximation \cite{Landau1987}. 
Thermo-osmotic slip is a creeping flow along the solid-fluid interface induced in the direction of temperature gradient \cite{Bregulla2016,Fraenzl2022,Xu2023,Tsuji2023} (see also Refs.~\cite{Fu2017,Ganti2017,Chen2021a,Herrero2022,Ouadfel2023,Ouadfel2024} for the results of molecular simulation). Thermo-osmotic slip can be modeled as a slip boundary condition with a slip coefficient, but the theoretical determination of a slip coefficient for a given set of fluid and surface properties is challenging\red{; readers are referred to as a classical attempt to obtain the thermal-slip boundary condition for liquids by Derjaguin et al. \cite{Derjaguin1987} (see also the comprehensive review paper \cite{Anderson1989}) and recent microscopic theory of thermo-osmosis \cite{Anzini2019,Anzini2022}.} 
Therefore, we set the slip coefficient a given constant in the present paper.  
Because of these thermally-induced flows, a thermo-fluid simulation is usually required to analyze the experimental results of photothermal heating. In fact, some of the references mentioned above carried out numerical simulations of temperature and/or flow fields based on a finite-element method using commercial software (mostly, COMSOL) 
\cite{Donner2011,Braun2013,FloresFlores2015,Bregulla2016,Kotnala2019a,ZentenoHernandez2020,Hayashi2021,Fraenzl2022,Zhou2023,Chen2024,Meng2024,Yang2023,Qian2020,Shi2023,Paul2022a,Jin2018,Weinert2009,Duhr2005}. 
Due to the localized heat generation, the characteristic lengths of temperature and flow fields can be multiscale, i.e., from the size of nanoscale localized heat to the size of millimeter-scale fluid channels. 
Inherently, some trial and error in mesh construction and/or long computational time is required. 
These difficulties prevent systematic parameter study on the temperature and flow fields and the search for optimized experimental conditions. 
Moreover, when not familiar with fluid mechanics, the construction of the code, the interpretation of numerical results, and their validation are time-consuming, being ``costly" not only in the sense of a license fee of commercial software. 

\begin{figure*}[bt]
    \centering
    \includegraphics[width=0.8\textwidth]{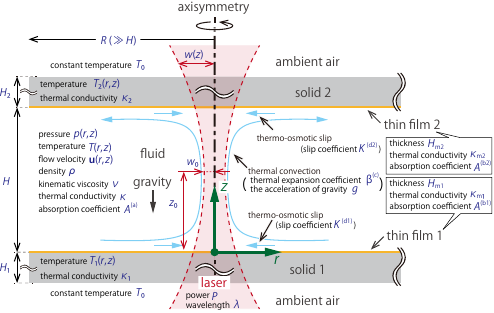}
    \caption{Schematic summary of the present problem and notations. A laser is irradiated to a fluid film between two parallel plates (i.e., substrates). The plates (solid 1 and 2 in the figure) are coated with light-absorbing thin films. The fluid and/or the thin films can be heated by the laser because of the photothermal effect. As a result, thermal convection and thermo-osmotic slip are induced in the channel. }
    \label{fig:problem}
\end{figure*}

A typical setting of photothermal experiment is a fluid film confined between two parallel plates, where a laser is irradiated in the perpendicular direction to the plates, as shown in Fig.~\ref{fig:problem}. To be more precise, the laser is typically a Gaussian beam that has a beam intensity profile described by a Gaussian function in the plane perpendicular to the beam axis (see Fig.~\ref{fig:problem}), and thus the resulting temperature and flow fields are safely assumed to be axisymmetric under microfluidic conditions such as millimeter-scale (or less) distance of the plates. 
For steady axisymmetric problems with the photothermal effect of Gaussian-type beams, an semi-analytical approach based on Hankel transformation is applicable, and temperature fields for a two-dimensional material \cite{Goushehgir2021} and for a laser without focusing effect (i.e., constant beam width in the beam propagation direction) \cite{Zablotsky2021} were obtained. (Unsteady temperature field was treated in Ref.~\cite{Peng2016}.) 
The present study tries to solve semi-analytically both temperature and flow fields for more general setting described in Fig.~\ref{fig:problem} which covers a wide range of steady cases introduced so far, and provide an easy-access instant computational tool (see the supporting information of this paper shared through Ref.~\cite{Tsuji2024a}); 
as an application of this developed tool, we also investigate the underlying mechanism of the optothermal trapping of nanoparticles observed in our previous paper \cite{Tsuji2021}. 
Here, the presence of the thin film is approximated by using an asymptotic expansion with respect to the film thickness, deriving the approximate boundary conditions at the liquid-solid interface. 
In addition to the assumption of the smallness of the film thickness, it should be noted for users of the developed computational tool that the limitation of the present result is the ignorance of (i) nonlinear effects, i.e., the convective terms in the heat-conduction and the Navier-Stokes equations and (ii) the temperature dependence of transport coefficients such as thermal conductivity and viscosity.

The structure of the paper is as follows. 
Section~\ref{sec:formula} introduces the statement of the problem, which can be decomposed into several linear systems.  
The semi-analytical and numerical methods for solving these linear systems are explained in Sec.~\ref{sec:method}. 
Then, the results are given in Sec.~\ref{sec:results} which is divided into two parts: first, the obtained semi-analytical and numerical solutions are validated in Sec.~\ref{sec:validations}. 
Next, the semi-analytical solution is applied to some typical situations of photothermal studies: thermal convection (Sec.~\ref{sec:convection}), thermo-osmotic slip (Sec.~\ref{sec:slip}), and optothermal trap (Sec.~\ref{sec:optothermal-trap}) to investigate their physical characteristics. 
Finally, the conclusion will be given in Sec.~\ref{sec:conclusion} followed by some future perspective.


\section{\label{sec:formula}Formulation}
We consider thermally-induced fluid flows where the temperature rise is triggered by the photothermal effects of (a) fluids and/or (b) surfaces of a fluid channel, as shown in Figs.~\ref{fig:heating}(a) and \ref{fig:heating}(b), respectively. 
Selective heating of the surface can be realized by, for instance, coating the channel wall by light-absorbing materials such as an Au thin film.
The temperature fields can induce fluid flows owing to (c) a buoyancy force due to gravity and/or (d) thermo-osmotic slip at the surfaces, as shown in Figs.~\ref{fig:heating}(c) and \ref{fig:heating}(d), respectively. 
The former is the so-called thermal convection or natural convection and is modeled through the Boussinesq approximation; the latter is phenomenologically modeled as a thermal-slip boundary condition with a slip coefficient. When the slip coefficient is negative, which is usual, the slip flow is directed from the colder to the hotter sides, as shown in Fig.~\ref{fig:heating}(d). 

In the present paper, we restrict ourselves to linear cases as detailed below, and thus two heating effects [Figs.~\ref{fig:heating}(a) and (b)] and resulting flows 
[Figs.~\ref{fig:heating}(c) and (d)] can be superposed. Therefore, a single formulation including all these effects will be presented, where quantities associated with (a) fluid heating, (b) surface heating, (c) thermal convection, and (d) thermo-osmotic slip will be denoted with superscript ``(a)", ``(b)", ``(c)", and ``(d)", respectively. For instance, the absorption coefficient of the fluid is denoted by $\Aa$.

\subsection{\label{sec:problem}Problem}
Let us consider a thin fluid film with a thickness of $H$ confined between two planar walls made of solid 1 and solid 2 with heights $H_1$ and $H_2$, respectively, as shown schematically in Fig.~\ref{fig:problem}. In most experiments, the solid parts (i.e., substrates) are thicker than the fluid film, i.e., $H_j>H$. 
The other faces of solid parts are in contact with stationary ambient air and are assumed to be kept at a constant temperature $T_0$. 
Thin films 1 and 2 with thickness $\Hm1$ and $\Hm2$, respectively, are attached to the inner surfaces of the corresponding solid parts. 
The film thickness $\Hmj$ $(j=1,\,2)$ is negligibly small compared to the characteristic length scales of the temperature and flow fields, e.g., the linear dimension of a heat source. Then, we can consider the thin films as ``surfaces" of the solid parts. 
The density, kinematic viscosity, thermal conductivity, and thermal expansion coefficient of the fluid are denoted by $\rho$, $\nu$, $\kappa$, and $\betac$, respectively, where $\betac$ has a superscript ``(c)" because it is the coefficient of the driving force of thermal convection (see also Fig.~\ref{fig:heating}). The thermal conductivities of the solid $j$ and the thin film $j$ $(j=1,\,2)$ are denoted by $\kj$ and $\kmj$, respectively. 

A focused Gaussian beam with a wavelength of $\lambda$ and power $P$ is irradiated to the above three-layer system (i.e., solid-fluid-solid layers) in the perpendicular direction to the surface, where the beam width $w$ is a function of the position $z$ in the direction of beam propagation. The beam width takes its minimum $w_0$ at a focal plane $z=z_0$. The acceleration of gravity $g$ is directed in the negative $z$ direction. The beam can heat up both fluid and surfaces (i.e., thin films), resulting in an inhomogeneous temperature field of fluid $T(r,z)$ and those of solids $T_j(r,z)$ $(j=1,\,2)$, where $r$ is a radial direction and an axisymmetric state is assumed. 
The absorption coefficients of the fluid and the thin film $j$ $(j=1,\,2)$ are denoted by $\Aa$ and $\Abj$, respectively.

The temperature fields can induce fluid flows $\v(r,z)=(v_r(r,z),v_z(r,z))$ and pressure field $p(r,z)$ due to thermal convection and/or thermo-osmotic slip at the surfaces, as shown in Fig.~\ref{fig:problem}. 
The slip coefficient at the surface $j$ is denoted by $\Kdj$ $(j=1,\,2)$.
The radial dimension $R$ of the whole system is assumed to be much larger than the vertical dimension $H$. Therefore, the flow and temperature variation vanish as $r\to\infty$. 

We investigate the steady state of temperatures $T$, $T_1$, $T_2$, and a flow field $\v$ under the following assumptions: 
\begin{enumerate}
\item The transport coefficients ($\kappa$, $\kj$, $\kmj$, $\nu$), the fluid density $\rho$, the thermal expansion coefficient $\betac$, the absorption coefficients ($\Aa$, $\Abj$), and the slip coefficient $\Kdj$ are all constant.  
\item The Reynolds number $\Rey=L_0 v_0/\nu$, where $L_0$ and $v_0$ are the characteristic length and speed, respectively, is so small that the nonlinear convection terms in the Navier-Stokes equation and the energy equation can be neglected. 
\end{enumerate}
These two assumptions allow us to linearize the whole system and to simplify the analysis so that an analytical approach is possible. 

\subsection{Basic equations}\label{sec:basic-equations}
Because we neglect the convection terms, the temperature fields are decoupled from the flow field. Therefore, we first present the energy equation for the temperatures.
The temperature of a fluid subject to a Gaussian beam and the temperatures of solids $j(=1,\,2)$ are described by steady heat-conduction equations: 
\begin{subequations}
\label{eq:temperature}
\begin{align}
&\kappa \left(\Dp T+\rnd{^2 T}{z^2}\right) + \Aa I(r,z) =0\quad (0< z<H), \label{eq:temperature-fluid}\\
& \k1 \left(\Dp T_1+\rnd{^2 T_1}{z^2}\right)=0 \quad (-H_1 < z < 0), \\
&\k2 \left(\Dp T_2+\rnd{^2 T_2}{z^2}\right)=0 \quad (H < z < H+H_2), 
\end{align}
\end{subequations}
where $\Dp$ is a short-hand notation of a differential operator: 
\begin{align}
\Dp \equiv 
\frac{1}{r}\left(\rnd{}{r}(r\rnd{}{r})\right). \label{eq:Dp} 
\end{align}
Here, $I(r,z)$ in Eq.~\eqref{eq:temperature-fluid} is the intensity distribution of a Gaussian beam defined as
\begin{subequations}\label{eq:laser}
\begin{align}
&I(r,z) = I_0 \frac{w_0^2}{w^2(z)} \exp\left(-\frac{2r^2}{w^2(z)}\right), \quad 
I_0 = \frac{2 P}{\pi w_0^2}, \label{eq:beam-intensity}\\
&w(z) = w_0 \sqrt{1+\frac{(z-z_0)^2}{\zR^2}}, \quad \zR = \frac{\pi w_0^2 }{\lambda}, \label{eq:beam-width}
\end{align}
\end{subequations}
where $I_0$ is a reference intensity and $\zR$ is the Rayleigh length. 
\red{The laser power $P$ is defined as $P=I_0 w_0^2$.}
The boundary conditions of Eq.~\eqref{eq:temperature} are 
usually the continuity conditions of the temperature fields and energy fluxes at the interfaces $z=0$ and $H$. However, because of the presence of heat-absorbing thin films, the boundary conditions are modified as
\begin{widetext}
\begin{subequations}
\label{eq:temperature-bc}
\begin{align}
&
\begin{cases}
T - T_1 = \dhf\dfrac{\Hm1}{\km1}
\left(\kappa\drnd{T}{z}+\k1\drnd{T_1}{z}\right), \\[1em]
\kappa\drnd{T}{z}-\k1\drnd{T_1}{z} = - \Ab1\Hm1 I - \dhf\km1\Hm1 \Dp (T + T_1), 
\end{cases}
(z=0), \\[1em]  
&
\begin{cases}
T_2 - T = \dhf\dfrac{\Hm2}{\km2}
\left(\k2\drnd{T_2}{z}+\kappa\drnd{T}{z}\right), \\[1em]
\k2\drnd{T_2}{z}-\kappa\drnd{T}{z} = - \Ab2 \Hm2 I - \dhf\km2\Hm2 \Dp (T_2 + T), 
\end{cases}
(z=H),  
\end{align}
\end{subequations}
\end{widetext}
the derivation of which are summarized in \SI~A. Note that setting $\Hm1=0$ (or $\Hm2=0$) recovers the usual continuity conditions of temperature and heat flux at the interface $z=0$ (or $z=H$). 
It should also be noted that the energy flux due to the work done by viscous stress is neglected here because it is a nonlinear effect.
The terms with beam intensity $I$ in Eq.~\eqref{eq:temperature-bc} represent the heat generation due to the photothermal effect of the thin films. 
The boundary conditions of $T_j$ at the outer faces are 
\begin{align}
T_1 = T_0 \quad (z=-H_1), \quad 
T_2 = T_0 \quad (z=H+H_2), \quad 
\label{eq:temperature-bc-z}
\end{align}
and the conditions in the radial direction are 
\begin{align}
\rnd{T}{r}=\rnd{T_J}{r} = 0\quad (r=0,\;r\to\infty), \label{eq:temperature-bc-r}
\end{align}
where the condition at $r=0$ is due to the axisymmetry.

The fluid motion is described by the Stokes equation under the Boussinesq approximation: 
\begin{subequations}\label{eq:Stokes}
\begin{align}
&\rnd{v_r}{r}+\frac{v_r}{r}+\rnd{v_z}{z}=0,\\
&
\frac{1}{\rho}\rnd{p}{r} =  \nu\left(\Dp v_r-\frac{v_r}{r^2}+\rnd{^2 v_r}{z^2}\right), \\
&
\frac{1}{\rho}\rnd{p}{z} = \nu\left(\Dp v_z + \rnd{^2 v_z}{z^2}\right) + g \betac(T-T_0), \label{eq:Stokes-z}
\end{align}
\end{subequations}
where the temperature $T$ enters in Eq.~\eqref{eq:Stokes-z} as an inhomogeneous term representing the buoyancy force. 
Boundary conditions for $\v=(v_r,\,v_z)$ are 
\begin{subequations}\label{eq:slip-bc}
\begin{align}
&v_r = -\Kd1 \rnd{T}{r}, \quad  v_z = 0 \quad (z=0), \\
&v_r = -\Kd2 \rnd{T}{r}, \quad  v_z = 0 \quad 
(z=H), 
\end{align}
\end{subequations}
where $\Kd1$ and $\Kd2$ are thermal-slip coefficients. 
The conditions in the radial direction are 
\begin{align}
\rnd{v_z}{r}=v_r=0\quad (r=0), \quad  
v_z=v_r=0\quad (r\to\infty),  
\label{eq:Stokes-bc-r}
\end{align}
where the condition at $r=0$ is due to axisymmetry. 

\red{It should be remarked that the macroscopic equations, i.e., the heat-conduction equations \eqref{eq:temperature} and the Stokes equation \eqref{eq:Stokes}, may not be valid at molecular scales. To be more specific, the Fourier law for the heat-conduction equation may not be valid in a thin film \cite{Majumdar1993} when the film thickness is comparable or smaller than the mean free path of phonons. In such cases, alternative molecular-scale frameworks should be required to analyze heat conduction \cite{Luo2013}. In liquids, the Navier-Stokes equations are invalid below nanometer length scale \cite{Bocquet2010,Kavokine2021}. In gases, when the characteristic length scale is comparable or smaller than the mean free path of gas molecules (about $70$~nm at atmospheric pressure), the Boltzmann equation is adequate to describe the gas behavior \cite{Sone2007} instead of the Navier-Stokes equations.}

In experiments, micro- and nanoparticles may be dispersed in the fluid. 
When passive particles are dispersed in the fluid, the forces that act on them can be expressed in a simple form, as described below. Here, the particles are assumed to be small and the suspension is dilute so that there is no inter-particle interaction and the presence of the particles does not modify Eqs.~\eqref{eq:temperature} and \eqref{eq:Stokes}. Moreover, the ineartia of particles is neglected, resulting in a quasi-steady motion where the Stokes drag is \red{opposed} by the thermophoretic force \cite{Wuerger2010}. 
Under these assumptions, the forces $\F=(F_r,\,F_z)$ acting on the particles \red{at rest} are expressed as
\begin{align}
\F = 3\pi d \eta (\v - D_T \nabla T),  \label{eq:force}
\end{align}
where $\eta=\rho \nu$ is the viscosity of the fluid, $d$ is the diameter of the dispersed particle, \red{$\v$ is the flow velocity}, and $D_T$ is the thermophoretic mobility of the particle. When $D_T$ is positive, the thermophoretic force, i.e., the second term on the right-hand side of Eq.~\eqref{eq:force}, is directed from hot to cold along the temperature gradient. 
It should also be mentioned that to obtain a more accurate expression of the force, we need to integrate the stress over the sphere surface, and Eq.~\eqref{eq:force} should be considered as a rough approximation. 
Moreover, thermophoretic particles may drive the flow around them (e.g., \red{Ref.}~\cite{Tsuji2023}) but the present model neglects it. 

Finally, we give a brief remark on the characterization of thermophoresis in experiments. 
It is difficult to separate the thermophoretic motion near the boundary from the particle transport caused by thermo-osmotic slip flows, because both motions are directed along a temperature gradient near the boundary. 
Therefore, \red{when} only the particle motion along the boundary is discussed, there is a risk of mistaking these two effects. For example, negative thermophoresis (from cold to hot) cannot be distinguished from particle transport caused by the usual thermo-osmotic slip (from cold to hot) near the boundary. 
Therefore, thermophoretic motion away from boundaries is preferable when focused only on thermophoresis, as discussed in Ref.~\cite{Tsuji2017}. 
Note that the case of negative $D_T$ is rare (see, e.g., \cite{Jiang2009,Chen2015c,Vigolo2010b,Tsuji2017,Tsuji2018b,Zhou2023,Kollipara2023}). Thus, when evaluating negative thermophoresis near the boundary, it is vital to examine carefully the effect of thermo-osmotic slip by changing physical parameters such as surface modification and/or the distance between particles and a substrate. 

\subsection{Non-dimensional expression} 
\begin{figure}[bt]
    \centering
    \includegraphics[width=\linewidth]{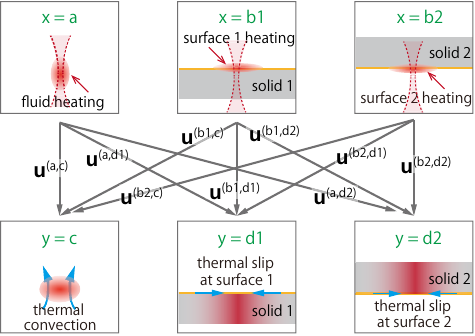}
    \caption{Schematic of decomposed problems.}
    \label{fig:decomposed}
\end{figure}

The system presented in Sec.~\ref{sec:basic-equations} includes three physical parameters $\Aa$, $\Ab1$, and $\Ab2$, i.e., the parameters characterizing the absorption of light into the fluid, surface 1, and surface 2, respectively. These absorption effects are directly related to the amount of heat generated there. Heating of the fluid, surface 1, and surface 2 are termed (a), (b1), and (b2) and induce inhomogeneous temperature fields, respectively. 
These temperature fields then cause three types of fluid flows, i.e., thermal convection and thermo-osmotic slip flows at surface 1 and surface 2, which are termed (c), (d1), and (d2), respectively. 
These effects of the $3\times 3$ types can be separated in the analysis due to the linearity of the system. In the following, we introduce non-dimensional parameters to systematically decompose the system. 
A schematic of the decomposition is illustrated in Fig.~\ref{fig:decomposed}. 

First, we introduce the magnitude of characteristic temperature variation (scaled by $T_0$) due to absorption: 
\begin{align}
\dTa = \frac{I_0 w_0^2 \Aa/\kappa}{ T_0} , \quad \dTbj = \frac{I_0 w_0^2  \Abj /\kappa}{ T_0} \quad (j=1,\,2). \label{eq:delta_T}
\end{align}
The beam waist $w_0$ is chosen as the characteristic length of temperature variation (i.e., the characteristic linear length of the heat source). Then, we introduce non-dimensional position variables $(\tr,\tz)$ through the relation $(r,z)=w_0(\tr,\tz)$. In the same manner, the non-dimensional thicknesses of the fluid $\tH$, the solid $\tHj$, and the thin film $\tHmj$ are defined by the relation $(H,H_j,H_{\m j})=w_0(\tH,\tHj,\tHmj)$. The beam radius $w(z)$ and the beam parameters are non-dimensionalized as $w(z)=w_0 \tw(\tz)$, $z_0=w_0\tz_0$, and $\zR=w_0\tzR/\sqrt{2}$. 
The thermal conductivity of the fluid, $\kappa$, is taken as a reference value, and the thermal conductivities of the solids and thin films are expressed as $\kj=\kappa\tkj$ and $\kmj=\kappa\tkmj$ $(j=1,\,2)$.

Using the non-dimensional parameters introduced in Eq.~\eqref{eq:delta_T}, the temperatures are decomposed as
\begin{widetext}
\begin{subequations}
\begin{align}
&T(r,z) = T_0( 1 + \tau(\tr,\tz)) = T_0\left(1+\dTa \ta\, (\tr,\tz) + \dTb1 \tb\,1 (\tr,\tz) + \dTb2 \tb\,2 (\tr,\tz)\right), \label{eq:temperature-decompose}\\
&T_1(r,z)=T_0(1 + \t1(\tr,\tz))=T_0\left(1+\dTa \ta1(\tr,\tz) + \dTb1 \tb11 (\tr,\tz) + \dTb2 \tb12(\tr,\tz)\right), \\
&T_2(r,z)=T_0(1 + \t2(\tr,\tz))=T_0\left(1+\dTa \ta2(\tr,\tz) + \dTb1 \tb21 (\tr,\tz) + \dTb2 \tb22(\tr,\tz)\right), 
\end{align}
\end{subequations}
(see also Fig.~\ref{fig:decomposed}) and then Eqs.~\eqref{eq:temperature}, \eqref{eq:temperature-bc}--\eqref{eq:temperature-bc-r} can be rearranged as
\begin{subequations}\label{eq:temperature-nd}
\begin{align}
&
\tDp \tx\, + \rnd{^2\tx\,}{\tz^2}  + \mIx=0
\quad \left(\tDp\equiv \frac{1}{\tr}\left(\rnd{}{\tr}(\tr\rnd{}{\tr})\right)\right),\\
&\tDp \tx1 + \rnd{^2\tx1}{\tz^2}=0, \quad 
\tDp \tx2 + \rnd{^2\tx2}{\tz^2}=0, \\
&
\begin{cases}
\tx\, - \tx1 = \dhf\dfrac{\tHm1}{\tkm1 }
\left(\drnd{\tx\,}{\tz}+\tk1\drnd{\tx1}{\tz}\right), \\[1em]
\drnd{\tx\,}{\tz}-\tk1\drnd{\tx1}{\tz} = - \Big(\mIx_1 + \dhf\tkm1 \tDp ( \tx\, + \tx1)\Big)\tHm1, 
\end{cases}
(\tz=0), \\[1em]  
&
\begin{cases}
\tx2 - \tx\, = \dhf\dfrac{\tHm2}{\tkm2}
\left(\tk2\drnd{\tx2}{\tz}+\drnd{\tx\,}{\tz}\right), \\[1em]
\tk2\drnd{\tx2}{\tz}-\drnd{\tx\,}{\tz} = - \Big( \mIx_2 + \dhf\tkm2 \tDp (\tx2 + \tx\,)\Big)\tHm2, 
\end{cases}
(\tz=\tH),  \\[1em] 
&\tx1 = 0\quad (\tz = -\tH_1), \quad 
\tx2 = 0 \quad (\tz = \tH + \tH_2), \\
&
\rnd{\tx\,}{\tr} =\rnd{\tx1}{\tr} = \rnd{\tx2}{\tr}  =0 \quad 
(\tr=0,\;\tr \to \infty), \label{eq:temperature-nd-bc-r}
\end{align}
\end{subequations}
\end{widetext}
where the superscript $(\x)$ is an indicator of the heat source, i.e., $\x = \a$ (fluid heating), $\b1$ (surface 1 heating), or $\b2$ (surface 2 heating); $\tx\,$ $\tx1$, and $\tx2$ are the non-dimensional temperature variation and
$\mIx$, $\mIx_1$, $\mIx_2$ $(\x = \a,\,\b1,\,\b2)$ are the inhomogeneous terms defined as
\begin{equation}
\begin{split}
&\mIa = \tI(\tr,\tz) \equiv \frac{1}{\tw^2(\tz)}\exp(-\frac{2\tr^2}{\tw^2(\tz)}), \\
&\tw(\tz) = \sqrt{1+\frac{2(\tz-\tz_0)^2}{\tzR^2}}, \\
&\mIb1_1 =  \tI(\tr,\,0), \quad 
\mIb2_2 =  \tI(\tr,\,\tH), \\
&\mIa_1 = \mIa_2=\mIb1=\mIb2=\mIb1_2=\mIb2_1=0. 
\label{eq:mIx}
\end{split}
\end{equation}

In a similar manner, we decompose the system for the fluid motion Eqs.~\eqref{eq:Stokes} and \eqref{eq:slip-bc}.  
First, we introduce non-dimensional parameters that characterize the magnitude of two types of fluid flows, i.e., thermal convection and thermo-osmotic slip flows, scaled by the reference speed $v_0=\nu/w_0$ as 
\begin{align}
\duc = \frac{g w_0^2 \betac T_0/\nu}{v_0}, \quad 
\dudj = \frac{\Kdj T_0/w_0}{v_0}
\quad (j=1,\,2). \label{eq:delta_u} 
\end{align}
Then, denoting by $p_0=\rho v_0^2$ a reference pressure, the flow velocity $\v(r,z)$ and the pressure $p(r,z)$ of the fluid can be decomposed as
\begin{widetext}
\begin{subequations}
\begin{align}
\v = v_0\u =  v_0
&\left( \dTa  \duc \uac +
 \dTa  \dud1 \uad1 +
 \dTa \dud2 \uad2 \right.\notag\\
&+
 \dTb1 \duc \ubc1 +
\dTb1 \dud1 \ubd11 +
\dTb1 \dud2 \ubd12 \notag\\
&+\left.
 \dTb2  \duc \ubc2 +
\dTb2 \dud1 \ubd21 +
\dTb2 \dud2 \ubd22 \right), \label{eq:Stokes-decompose}\\
p = p_0 \tp = p_0
&\left(\dTa \duc \tpac +
 \dTa\dud1 \tpad1 +
 \dTa \dud2 \tpad2 \right.\notag\\
&+
\dTb1 \duc \tpbc1 +
\dTb1 \dud1 \tpbd11 +
\dTb1 \dud2 \tpbd12 \notag\\
&+\left.
\dTb2 \duc \tpbc2 +
\dTb2 \dud1 \tpbd21 +
\dTb2 \dud2 \tpbd22 \right), 
\end{align}
\end{subequations}
\end{widetext}
where $\uxy(\tr,\tz)$ and $\tpxy(\tr,\tz)$ with the superscripts $\x=\a$, $\b 1$, $\b2$ and $\y=\c$, $\dd 1$, $\dd 2$ are the non-dimensional flow velocities and the non-dimensional pressures, respectively, with the superscript $\y$ being an indicator of induced flows. For instance, $\uac$ is (c) the thermal convection induced by (a) the heating of the fluid; $\ubd12$ is (d) the thermo-osmotic slip at surface 2 induced by (b) the heating of the surface 1 (see Fig.~\ref{fig:decomposed}). Now, Eqs.~\eqref{eq:Stokes}--\eqref{eq:Stokes-bc-r} are rearranged as 
\begin{subequations}\label{eq:Stokes-nd}
\begin{align}
&\rnd{\urxy}{\tr}+\frac{\urxy}{\tr}+\rnd{\uzxy}{\tz}=0,\\
&
\rnd{\tpxy}{\tr} =  \tDp \urxy-\frac{\urxy}{\tr^2}+\rnd{^2 \urxy}{\tz^2}, \\
&
\rnd{\tpxy}{\tz} = \tDp \uzxy + \rnd{^2 \uzxy}{\tz^2} + \mJxy, \\[0.5em]
&\begin{cases}
\urxy =  \mJxy_1, \quad  \uzxy = 0 \quad (\tz=0), \\[0.5em]
\urxy =  \mJxy_2, \quad  \uzxy = 0 \quad (\tz=\tH), 
\end{cases}\\[0.5em]
&
\rnd{\uzxy}{\tr}=\urxy=0\quad (\tr=0), \\
&
\uzxy=\urxy=0\quad (\tr\to\infty), \quad \label{eq:Stokes-nd-bc-r}
\end{align}
\end{subequations}
where the inhomogeneous terms are defined as
\begin{equation}
\begin{split}
&\mJxc = \tx\,, \quad 
\mJxd1_1 = -\drnd{\tx\,}{\tr}, \quad 
\mJxd2_2 = -\drnd{\tx\,}{\tr}, \\
&\mJxd1 = \mJxd2 = 0, \\
&\mJxc_1 = \mJxd2_1 = \mJxc_2 = \mJxd1_2 = 0, 
\end{split}
\end{equation}
with $\x=\a,\,\b1,\,\b2$. 
\red{Note that thermal convection is usually investigated in terms of the non-dimensional parameters such as the Rayleigh number and/or the Grashof number. Here, the Grashof number (based on the channel height $H$) can be expressed by $\mathit{Gr}=(g \betac H^3/\nu^2)\Delta T$, with $\Delta T$ a characteristic temperature increase. Since $\Delta T \approx \max_\x\dTx T_0$, where $\max$ is taken for $\x\,=\,\a,\,\b1,\,\b2$, we then have $\mathit{Gr}\approx \max_\x\dTx\duc (H/w_0)^3$. The Rayleigh number is then defined as $\mathit{Ra}=\mathit{Gr}\mathit{Pr}$ with $\mathit{Pr}$ the Prandtl number, which is about $\mathit{Pr}=7$ for water. These observation will be used later in Sec.~\ref{sec:convection} when discussing thermal convection.}

The systems Eqs.~\eqref{eq:temperature-nd} and \eqref{eq:Stokes-nd} with the beam intensity Eq.~\eqref{eq:mIx} include the physical parameters
\begin{align}
\tz_0,\quad 
\tzR, \quad 
\tH, \quad 
\tHj, \quad 
\tkj, \quad 
\tHmj, \quad  
\tkmj \quad 
(j=1,\,2). \label{eq:parameters}
\end{align}
These systems are solved semi-analytically in Sec.~\ref{sec:semi-anal} and numerically in Sec.~\ref{sec:nume-anal}. 
Note that the beam intensity $I_0$, the acceleration of gravity $g$, the absorption coefficients $\Aa$ and $\Abj$, the thermal expansion coefficient $\betac$, and the slip coefficient $\Kdj$ are all included in $\dTx$ (Eq.~\eqref{eq:delta_T}) and $\duy$ (Eq.~\eqref{eq:delta_u}) and excluded from the systems Eqs.~\eqref{eq:temperature-nd} and \eqref{eq:Stokes-nd}. 

Finally, non-dimensional force $\f$ acting on the dispersed particles is defined as (see also Eq.~\eqref{eq:force})
\begin{align}
\F = 3\pi d \eta v_0 \f = 3\pi d \eta v_0 \left( \u - \tD_T \tnabla \tau\right), \; 
\tD_T = \frac{T_0}{w_0 v_0}D_T, \label{eq:force-nd}
\end{align}
where $\tD_T$ is a non-dimensional thermophoretic mobility and $\tnabla$ is a gradient operator in the $(\tr,\tz)$ space. 


\section{Method of analysis}\label{sec:method}
\subsection{Semi-analytical solutions}\label{sec:semi-anal}
The derivation of the semi-analytical solutions below is based on the Hankel transform and are described in \SI~B. Here, only the overview of the results is given. 

The solutions of Eqs.~\eqref{eq:temperature-nd} and \eqref{eq:Stokes-nd} for the fluid, namely, $\tx\,$, $\urxy$, and $\uzxy$, together with related derivatives used in Eq.~\eqref{eq:force-nd}, can be expressed as, for $\x=\a,\,\b1,\,\b2$ and $\y=\c,\dd 1,\,\dd 2$, 
\begin{align}
\begin{bmatrix}
\tx\,\\
\partial\tx\,/\partial \tr \\
\partial\tx\,/\partial \tz \\
\urxy \\
\uzxy 
\end{bmatrix}
 = \bigintss_0^\infty s 
\begin{bmatrix}
  \btx\,(s,\tz) J_0(s\tr)\\
-s\btx\,(s,\tz) J_1(s\tr)\\
(\partial\btx\,/\partial \tz)J_0(s\tr) \\
\burxy(s,\tz)J_1(s\tr) \\
\buzxy(s,\tz)J_0(s\tr) 
\end{bmatrix}
\dd s, \label{eq:solution}
\end{align}
where $J_0$ and $J_1$ are the Bessel functions of the order 0 and 1, respectively, and  $\btx\,(s,\tz)$,  $\burxy(s,\tz)$, and $\buzxy(s,\tz)$ in the right-hand side are the Hankel transforms of $\tx\,$, $\urxy$, and $\uzxy$, respectively; they are explicit functions of $(s,\tz)$ though lengthy and are given in \SI~C. Therefore, $\tx\,$, $\urxy$, and $\uzxy$ are semi-explicit: a single numerical integration yields the solutions for each decomposed problem (Eqs.~\eqref{eq:temperature-nd} and \eqref{eq:Stokes-nd}). The solution to the full problem Eqs.~\eqref{eq:temperature}--\eqref{eq:Stokes-bc-r} can be obtained by the weighted superposition (i.e., Eqs.~\eqref{eq:temperature-decompose} and \eqref{eq:Stokes-decompose}) of $\tx\,$, $\urxy$, and $\uzxy$ with weights $\dTx$ (Eq.~\eqref{eq:delta_T}) and $\duy$ (Eq.~\eqref{eq:delta_u}). Other quantities such as $\tx1$, $\tx2$, and $\tpxy$ can be treated in the same manner as $\tx\,$ but the details are omitted here (see \SI~B for more details). 

Because there is no iterative procedure, which is usually required in the numerical simulation of partial differential equations, the computation of Eq.~\eqref{eq:solution} is practically instant. For instance, a couple of seconds is required for the computation of the entire domain with descent grid points of $O(10^4)$ using a standard laptop. A Python or Fortran code to compute the temperature and flow fields is freely available in the supporting information with a user guide \cite{Tsuji2024a}. 

\subsection{Numerical analysis}\label{sec:nume-anal}

We carry out the numerical analysis of the problem shown in Sec.~\ref{sec:basic-equations} and Fig.~\ref{fig:problem}. 
The detail of a numerical scheme and its accuracy check is presented in \SI~D and we briefly state here the motivation and overview of the numerical analysis. 
In short, the role of the numerical analysis is two-fold: 
(i) the validation of the thin-film approximation Eq.~\eqref{eq:temperature-bc} (see also \SI~A) and (ii) the validation of the semi-analytical solution Eq.~\eqref{eq:solution}.

The point (i) is done by comparing two numerical solutions. One is the numerical solution of the temperature (i.e., the solution of Eqs.~\eqref{eq:temperature}--\eqref{eq:temperature-bc-r}) with the boundary condition Eq.~\eqref{eq:temperature-bc} using the thin-film approximation. We call this system the three-layer system. The other is the numerical solution of the temperature in a five-layer system shown in Fig.~\ref{fig:problem} without the thin-film approximation. Note that the boundary conditions without the thin-film approximation mean that the right-hand sides in Eq.~\eqref{eq:temperature-bc} all vanish (e.g., $T=T_1$ at $z=0$). 
In the five-layer system, in addition to the three-layer system (Eqs.~\eqref{eq:temperature}--\eqref{eq:temperature-bc-r}), we solve numerically the heat-conduction equations inside the thin films of thicknesses $\Hm1$ and $\Hm2$ together with the continuities of the temperature and the heat fluxes in the surface-normal direction at the interfaces. We omit the governing equations for the temperatures in the thin films because they are essentially the same as those for the fluid and the solids except for the thermal conductivities. 
The point (ii) is validated by comparing the numerical solution of the three-layer system and the semi-analytical solution Eq.~\eqref{eq:solution}. 

We use a standard finite-difference method with a second-order central difference scheme to obtain the temperature fields (see Eq.~\eqref{eq:temperature}). 
The steady state is obtained by an iteration procedure using the Gauss-Seidel method. 
To obtain the flow velocity field, Eq.~\eqref{eq:Stokes} is solved based on a stream function-vorticity formulation. That is, the Stokes equation Eq.~\eqref{eq:Stokes} is converted into the coupled partial differential equations of the vorticity $\omega_\ast=\rnd{v_r}{z}-\rnd{v_z}{r}$ and the stream function $\psi_\ast$ that satisfies the relation $(v_r,v_z)=(\frac{1}{r}\rnd{\psi_\ast}{z},-\frac{1}{r}\rnd{\psi_\ast}{r})$. Note that the equation for vorticity incorporates the source terms due to the temperature variations, as shown in \SI~D. 
We use finite-difference methods similar to the temperature solver to obtain the numerical solutions of $\psi_\ast$ and $\omega_\ast$. 

\begin{table*}[tb]
\centering
\def\valkap {$0.5$}  
\def\valkapa{$1.0$}
\def\valkapb{$1.0$}
\def\valH   {$40$}
\def\valHa  {$80$}
\def\valHb  {$80$}
\def\valw   {$10$}
\def\valz   {$0$}
\def\valP   {$5\times10^{-2}$}
\def\vallamb{$1480$}
\def\comkap {TC of fluid  }
\def\comkapa{TC of solid 2}
\def\comkapb{TC of solid 1}
\def\comH   {height of fluid film        }
\def\comHa  {height of solid 1           }
\def\comHb  {height of solid 2           }
\def\comw   {beam width at FP  }
\def\comz   {$z$ position of FP}
\def\comP   {laser power                    }
\def\comlamb{wavelength of the laser        }
\def\valbec{$2.4\times10^{-4}$}
\def\valKda{$-2$}
\def\valKdb{$-2$}
\def\valrho{$1\times10^3$}
\def\valnu {$1\times10^{-6}$}
\def\valg  {$9.8$}
\def\valT  {$298$}
\def\combec{thermal expansion coef.}
\def\comKda{slip coef. at surface 1}
\def\comKdb{slip coef. at surface 2}
\def\comrho{fluid density}
\def\comnu {kinematic viscosity}
\def\comg  {gravity acceleration}
\def\comT  {reference temperature}
\def\valkmaA{0}
\def\valHmaA{0}
\def\valHmbA{0}
\def\valkmaB{1.0}
\def\valHmaB{0.5}
\def\valHmbB{0}
\def\valAaA{$2.4\times10^3$}
\def\valAbaA{0}
\def\valAbbA{0}
\def\valAaB{0}
\def\valAbaB{$1\times 10^4$}
\def\valAbbB{0}
\def\comkma{TC of thin film 1}
\def\comHma{thickness of thin film 1}
\def\comHmb{thickness of thin film 2}
\def\comAa{AC of fluid}
\def\comAba{AC of thin film 1}
\def\comAbb{AC of thin film 2}
\caption{Physical parameters used in Sec.~\ref{sec:validations}. 
(TC) thermal conductivity; (FP) focal plane; (AC) absorption coefficient}
\label{tab:phys-param}
{\tabcolsep = 0.5em
\small
\begin{tabular}{llllllll}
\hline 
\multicolumn{7}{l}{parameters common to cases A and B}\\[0.5em]
$\kappa$~[W/(m\;K)]      &\valkap &\comkap & &$w_0$~[\um]      &\valw   &\comw   \\
$\kappa_1$~[W/(m\;K)]    &\valkapa&\comkapa& &$z_0$~[\um]      &\valz   &\comz   \\
$\kappa_2$~[W/(m\;K)]    &\valkapb&\comkapb& &$P$~[W]          &\valP   &\comP   \\
$H$~[\um]                &\valH   &\comH   & &$\lambda$~[nm]   &\vallamb&\comlamb\\
$H_1$~[\um]              &\valHa  &\comHa  & &$\nu$~[m$^2$/s]  &\valnu  &\comnu  \\
$H_2$~[\um]              &\valHb  &\comHb  & &$g$  ~[m/s$^2$]  &\valg   &\comg   \\
$\betac$~[1/K]           &\valbec &\combec & &$T_0$~[K]        &\valT   &\comT   \\
$\Kd1$~[\um$^2$/(s\;K)]  &\valKda &\comKda & &$\rho$~[kg/m$^3$]&\valrho &\comrho \\
$\Kd2$~[\um$^2$/(s\;K)]  &\valKdb &\comKdb & &                 &        &        \\
\hline
\end{tabular}
\\[1em]
\begin{tabular}{lllllllll}
\hline 
\multicolumn{8}{l}{parameters different in cases A and B}\\[0.5em]
& \multicolumn{2}{c}{case} & & && \multicolumn{2}{c}{case} &\\
\cline{2-3}\cline{7-8}
& A & B & & && A & B &\\
$\km1$~[W/(m\;K)] &\valkmaA&\valkmaB&\comkma& &$\Aa$~[1/m]  &\valAaA &\valAaB &\comAa \\
$\Hm1$~[\um]      &\valHmaA&\valHmaB&\comHma& &$\Ab1$~[1/m] &\valAbaA&\valAbaB&\comAba\\
$\Hm2$~[\um]      &\valHmbA&\valHmbB&\comHmb& &$\Ab2$~[1/m] &\valAbbA&\valAbbB&\comAbb\\
\hline
\end{tabular}
}
\end{table*}


\section{\label{sec:results}Results and Discussion}

\subsection{Some validations using the numerical solutions}\label{sec:validations}
\subsubsection{Validation of the semi-analytical solution}

\begin{figure}[bt]
    \centering
    \includegraphics[width=\linewidth]{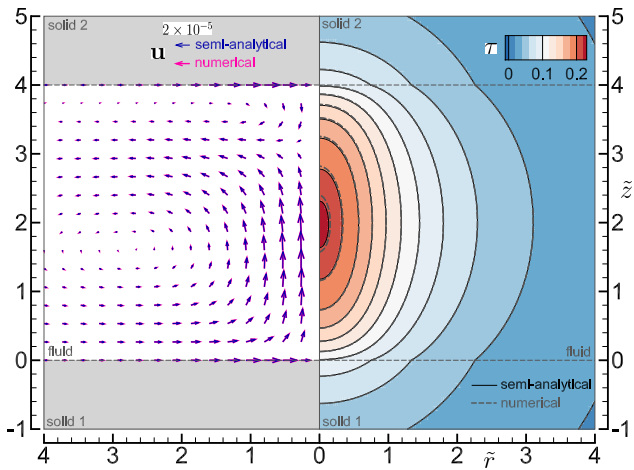}
        \caption{Case A: fluid heating (see Table~\ref{tab:phys-param}). (left) flow velocity field $\u$ for semi-analytical solution (blue arrows) and numerical  solution (pink arrows). (right) temperature field $\tau$ for semi-analytical solution (solid contours and heatmap) and numerical solutions (dashed contours). Note that no thin film is present in case A. }
    \label{fig:bh-ana-vs-num}
\end{figure}
\begin{figure}[bt]
    \centering
    \includegraphics[width=\linewidth]{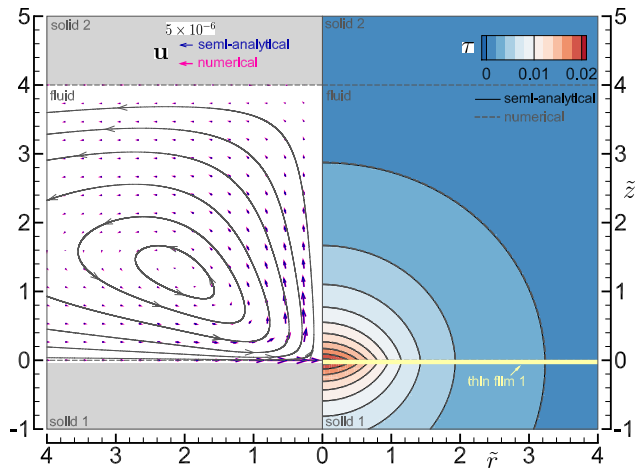}
    \caption{Case B: surface heating (see Table~\ref{tab:phys-param}). (left) flow velocity field $\u$ for semi-analytical solution (blue arrows) and numerical solution (pink arrows). Streamlines are also shown to visualize the flow in the region where $|\u|$ is small. (right) temperature field $\tau$ for semi-analytical solution (solid contours and heatmap) and numerical solutions (dashed contours). Note that only the thin film 1 is present in case B, which is shown near $\tz=0$ with yellow color.}
    \label{fig:th-ana-vs-num}
\end{figure}

We compare semi-analytical and numerical solutions using physical parameters in Table~\ref{tab:phys-param}, i.e., cases A (fluid heating) and B (surface heating). For the numerical solution, the finest grid parameters A1 and B1 in Table~S4 of \SI~D are used. 

Figure~\ref{fig:bh-ana-vs-num} shows the result of case A (fluid heating). The left and right halves show the flow velocity vectors and the temperature contours, respectively. Note that $\tr = 0$ is the axis of symmetry. The semi-analytical (or numerical) solution is shown using blue (or pink) vectors and solid (or dashed) contours. 
First, we describe the flow behaviors in Fig.~\ref{fig:bh-ana-vs-num}. It is seen that the fluid is heated and so are the solids due to the heat conduction. Concerning the flow field, the overall thermal convection in the counter-clockwise direction occurs due to the elevated temperature. In addition to the convection flow, thermo-osmotic slip flows near the interfaces at $\tz=0$ and $\tz=4$ in the negative $\tr$ direction are also induced. These slip flows enhance (or cancel out) the thermal convection near $\tz =0$ (or $\tz = 4$). There is little difference between the semi-analytical and numerical solutions; one may notice a slight difference in the temperature field near $(\tr,\tz)=(0,2)$. 

Next, we describe the flow behaviors in Fig.~\ref{fig:th-ana-vs-num} for case B (surface heating), where the legend of the figure is basically the same as Fig.~\ref{fig:bh-ana-vs-num}. Streamlines are added here to facilitate grasping the overall flow structure. The fluid (and the solid 1) is heated near $(\tr,\tz)=(0,0)$ in the thin film, although the magnitude of the temperature increase is milder than that of case A. Thermal convection in the counterclockwise direction is strongly enhanced near $\tz=0$ through a significant thermo-osmotic slip flow induced by the localized temperature increase. At $\tz=4$, the temperature variation is almost absent and so is the flow. As in the case of  Fig.~\ref{fig:bh-ana-vs-num}, there is little difference between the semi-analytical and numerical solutions.

To compare the semi-analytical and numerical solutions in Figs.~\ref{fig:bh-ana-vs-num} and \ref{fig:th-ana-vs-num}, we define their relative difference $E_{\num-\ana}(X)$ $(X=\tau,\,\u)$ as 
$ E_{\num-\ana}(X) = |X_{\num}-X_{\ana}|/\max |X_{\ana}|$, where $\max$ in the denominator is taken over whole grid points in the fluid domain and the subscript ``$\num$" and ``$\ana$" indicate the numerical and semi-analytical solutions, respectively. 
The temperature and flow velocity show the differences $E_{\num-\ana}(\tau)=0.7~\%$ and $E_{\num-\ana}(\bm{u})=1.5~\%$ in case A (fluid heating); 
$E_{\num-\ana}(\tau)=0.7~\%$ and $E_{\num-\ana}(\bm{u})=0.8~\%$ in case B (surface heating) at most. 
The difference is caused by the numerical error in the numerical solution, because improving the accuracy of the numerical integration of the semi-analytical solution Eq.~\eqref{eq:solution} does not change these values while the numerical solution is affected by the accuracy (i.e., the number of the grid points) as separately investigated in \SI~D. From the comparison, we conclude that both the semi-analytical and numerical scheme are validated.  


\subsubsection{Validation of the thin-film approximation}

\begin{figure*}[bt]
    \centering
    \includegraphics[width=0.6\linewidth]{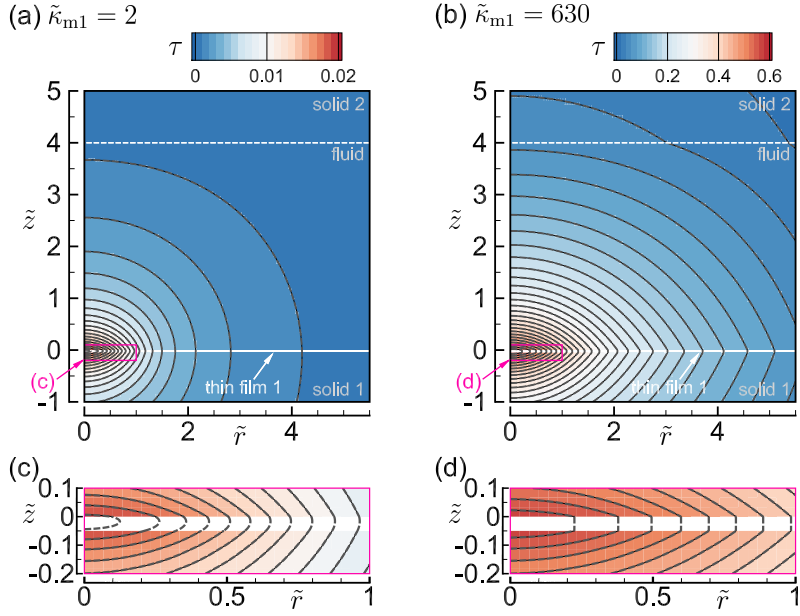}
    \caption{Temperature field $\tau$ in case B (surface heating; see Table~\ref{tab:phys-param}) with (a,c) $\tkm1=2$ and (b,d) $\tkm1=630$. The results of the three-layer system with the thin-film approximation (solid contour and heatmap) and those of the five-layer system without the thin-film approximation (dashed contours) are overlaid. (a,b) overview and (c,d) the magnification inside the thin film 1. }
    \label{fig:3L-vs-5L_overview}
\end{figure*}
\begin{figure*}[bt]
    \centering
    \includegraphics[width=0.8\linewidth]{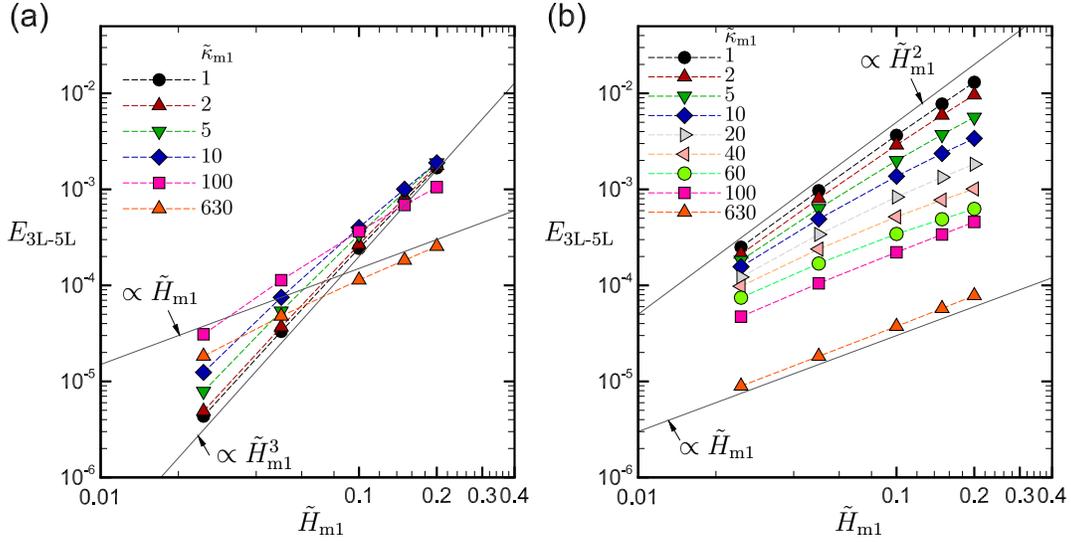}
    \caption{Relative difference $E_{\text{3L-5L}}$ as a function of the (non-dimensional) film thickness $\tHm1$ for various values of thermal conductivity $\tkm1$ of the thin film in (a) case A (fluid heating) and (b) case B (surface heating). Other physical parameters are set according to Table~\ref{tab:phys-param}. The lattice systems for this figure are coarse ones: we use A3 and B3 for three-layer systems and ALc and BLc for five-layer systems in Table~S4.}
    \label{fig:Hm1_dependence}
\end{figure*}

We compare the numerical solutions of the three-layer system and the five-layer system to validate the thin-film approximation Eq.~\eqref{eq:temperature-bc} (see also \SI~A). Note that the temperature field inside the thin film is also numerically computed in the five-layer system. Here, we show only the temperature field because the thin-film approximation explicitly applies to the temperature field. 
We use the grid parameters A1 and B1 (or AL and BL) for the three-layer system (or the five-layer system) in Table~S4 in \SI~D. 

Figure~\ref{fig:3L-vs-5L_overview} shows the temperature field $\tau$ for case B (Table~\ref{tab:phys-param}; surface heating) with different $\km1$, i.e., the thermal conductivity of the thin film. Figures~\ref{fig:3L-vs-5L_overview}(a) and \ref{fig:3L-vs-5L_overview}(b) are the results of (a) $\tkm1=\km1/\kappa=2$ [assume $\km1=1$ W/(m K) for e.g., polymer] and (b) $\tkm1=\km1/\kappa=630$ [assume $\km1=315$ W/(m K) for metal such as gold], respectively, where $\kappa=0.5$ W/(m K). Solid contours with heatmap (or dashed contours) indicate the results of the three-layer (or the five-layer) system. 
Figures~\ref{fig:3L-vs-5L_overview}(c) and \ref{fig:3L-vs-5L_overview}(d) are the magnification inside the thin film of Figs.~\ref{fig:3L-vs-5L_overview}(a) and \ref{fig:3L-vs-5L_overview}(b), respectively. A white region $(-\tHm1=-0.05\leq \tz\leq 0)$ indicates the thin film, and only the data of the five-layer system are available. 
It is seen that $\tau$ increases near the heat source at $\tr\approx 0$ inside the thin film $(-\tHm1\leq \tz\leq 0;\;\tHm1=0.05)$, and the generated heat spreads over the fluid and solid domains. The case of high thermal conductivity $\tkm1$ [Fig.~\ref{fig:3L-vs-5L_overview}(b)] results in larger temperature increase than that of low thermal conductivity [Fig.~\ref{fig:3L-vs-5L_overview}(a)]. The magnification [Figs.~\ref{fig:3L-vs-5L_overview}(c,d)] shows that the solid contours (the three-layer system) are almost identical with the dashed contours (the five-layer system) even though the former is obtained without solving the interior of the thin film.
From the figure, we can conclude that the three-layer system reproduces well the results of the five-layer system. 

To compare the numerical solutions of the three-layer system and the five-layer system quantitatively, we define their relative difference $E_{\text{3L-5L}}$ as 
$E_{\text{3L-5L}} = 
\displaystyle\max(|\tau_{\text{3L}}-\tau_{\text{5L}}|/ |\tau_{\text{3L}}|)$, 
where $\max$ is taken over whole grid points in the fluid domain and the subscripts ``$\text{3L}$" and ``$\text{5L}$" indicate the temperatures of the three- and five-layer systems, respectively. 
The quantity $E_{\text{3L-5L}}$ indicates a relative error due to the introduction of the thin-film approximation.

Figure~\ref{fig:Hm1_dependence}(a) shows the effect of the film thicknesses $\tHm1$ on $E_{\text{3L-5L}}$ for various thermal conductivities $\tkm1$, while the other physical parameters are set according to the case A (fluid heating) in Table~\ref{tab:phys-param}. 
For moderate values of $\tkm1$, the error $E_{\text{3L-5L}}$ is proportional to $\tHm1^3$, as predicted by the asymptotic expansion in \SI~A (see, e.g., Eq.~(A18) with $\ep=\tHm1=\Hm1/w_0$). For large $\tkm1$, $E_{\text{3L-5L}}\propto\tHm1$, which is worse than $\tHm1^3$, but still the error can be considered negligible: $\tHm1(=\ep)=0.1$ results in the error $E_{\text{3L-5L}}<0.1$~\%.
Figure~\ref{fig:Hm1_dependence}(b) shows the corresponding analysis for the case B (surface heating) in Table~\ref{tab:phys-param}. In this case, $\tau$ itself is proportional to $\tHm1$ as predicted by the semi-analytical solution Eq.~(C1a) and (C2)  with $\x=\b1$. Therefore, the relative error $E_{\text{3L-5L}}$ should be proportional to $\tHm1^2$, which is realized in Fig.~\ref{fig:Hm1_dependence}(b) for moderate $\tkm1$. As $\tkm1$ becomes large, the behavior $E_{\text{3L-5L}}\propto\tHm1^2$ transits to $E_{\text{3L-5L}}\propto\tHm1$ as in the case of panel (a). 
Nonetheless, the error can be considered negligible: $\tHm1(=\ep)=0.1$ results in the error $E_{\text{3L-5L}}<1$~\% for a wide range of $\tkm1$. 
Through this comparison, we conclude that the validity of the thin-film approximation is quantitatively confirmed. 


\subsection{Applications of the semi-analytical solutions}\label{sec:application}

In the following, we present the semi-analytical solutions obtained using the open source \cite{Tsuji2024a} developed in this study. That is, all the results can be reproduced by anyone with simple steps without any computational burden. 

In this section, we present dimensional quantities to obtain intuition from experimental view points. Note that the temperature increase and the resulting flow speed are linear functions of some physical parameters because of the linearization assumption [see also Eqs.~\eqref{eq:delta_T}, \eqref{eq:temperature-decompose} for temperature and Eqs.~\eqref{eq:delta_u}, \eqref{eq:Stokes-decompose} for velocity]. 
To be more precise, the temperature increase and the resulting flow speed depend linearly on the laser power $P(=I_0 w_0^2)$, the absorption coefficients $\Aa$ (and $\Abj$), the gravity acceleration $g$, the thermal expansion coefficient $\betac$, the slip coefficients $\Kdj$ $(j=1,\,2)$, and $\nu^{-1}$.
On the other hand, $w_0$ and $\kappa$ are used as reference quantities. That is, they affect the physical parameters Eq.~\eqref{eq:parameters}, which enter the semi-analytical solution in a complicated manner as shown in \SI~C. Therefore, the effects of beam width $w_0$ and thermal conductivity $\kappa$ on the flow behavior are not apparent and thus will be investigated below. 

\begin{figure*}[bt]
    \centering
    \includegraphics[width=0.7\linewidth]{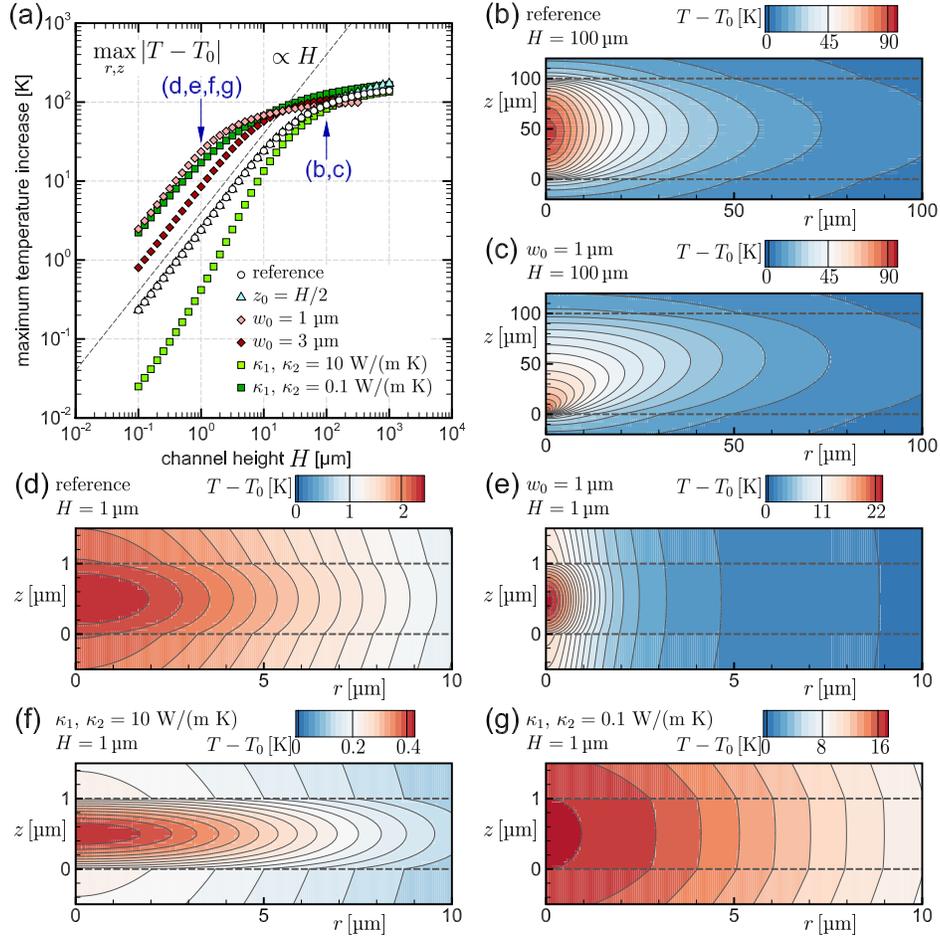}
    \caption{(a) Maximum temperature increase for various physical parameters, where unmentioned parameters are set to case A in Table~\ref{tab:phys-param}. (b-g) Temperature fields for some selected physical parameters in panel (a); note that the different length scales are taken for the $r$ and $z$ axes to grasp overall features.}
    \label{fig:convection-T}
\end{figure*}

\subsubsection{Thermal convection induced by fluid heating}\label{sec:convection}

In the photothermal experiments of microfluidics, the effect of thermal convection is inevitable under the presence of gravity. Here, we demonstrate the convenience of the present semi-analytical solution (see Eq.~\eqref{eq:solution} in Sec.~\ref{sec:semi-anal}) focusing on the characteristics of thermal convection over various physical parameters. 

In this section, we use the parameter set case A in Table~\ref{tab:phys-param} unless otherwise stated. The ``reference" case (white circles) in Fig.~\ref{fig:convection-T}(a) shows the dependence of the channel height $H$ in the range from $0.1$~\um~to $1$~mm on the maximum temperature increase $\max|T-T_0|$, where $\max$ is taken over $(r,\,z)$ in the whole fluid domain. 
\red{It should be noted that the cases with  $H=0.1$~\um~and $1$~mm corresponds to the Rayleigh number $\mathit{Ra}=4\times10^{-9}$ and $\mathit{Ra}=4\times 10^{3}$, respectively.} In the same figure, we also show the cases of the focal plane $z_0=H/2$ (sky-blue triangles), the beam width $w_0=1$~\textmu m (orange diamonds), the beam width $w_0=3$~\textmu m (red diamonds), and the thermal conductivities $\k1=\k2=10$ W/(m K) (light-green squares) and $\k1=\k2=0.1$ W/(m K) (green squares). 
Figures~\ref{fig:convection-T}(b-g) show the corresponding temperature fields for some selected parameters: (b,\,c) $H=100$~\textmu m and (d--g) $H=1$~\textmu m; (b,\,d) reference, (c,\,e) $w_0=1$~\textmu m, (f) $\k1=\k2=10$ W/(m K), (g) $\k1=\k2=0.1$ W/(m K).

Figure~\ref{fig:convection-T}(a) shows that the temperature increase is proportional to $H$ when $H$ is comparable or smaller than $w_0$ [see, reference ($w_0=10$~\textmu m), $w_0=1$~\textmu m, and $w_0=3$~\textmu m], and tends to saturate as $H$ becomes larger. The position of the focal plane $z_0$ does not effectively influence the temperature increase (see, reference and $z_0=H/2$). Smaller $w_0$ leads to higher temperature increase; smaller (or larger) thermal conductivities of the solids (i.e., channel walls) yield larger (or smaller) temperature increase. Note that the temperatures at large $H$ exceed the boiling point of water, indicating that the laser power $P=50$~mW is too strong for experiments using water in a microchannel with moderate $H$ (say, $H=100$~\textmu m). 
Figures~\ref{fig:convection-T}(b,\,c) and (d,\,e) show the effect of the beam width $w_0$ on the temperature profile for large $H$ and small $H$, respectively, and it is seen that using the smaller $w_0$ can localize and intensify the temperature increase. Figures~\ref{fig:convection-T}(d,\,f,\,g) show the effect of the thermal conductivity of the solid parts. Large/small thermal conductivity $\kj$ $(j=1,\,2)$ decreases/increases the temperature \red{variation}. It is also seen that the temperature contours are more aligned horizontally near $r=0$ in the case of large $\kj$ [Fig.~\ref{fig:convection-T}(f)], indicating the weak temperature gradient in the $r$ direction. On the other hand, the case of small $\kj$ [Fig.~\ref{fig:convection-T}(g)] results in the temperature contours more aligned vertically, indicating the weak temperature gradient in the $z$ direction. 

\begin{figure*}[tb]
    \centering
    \includegraphics[width=0.7\linewidth]{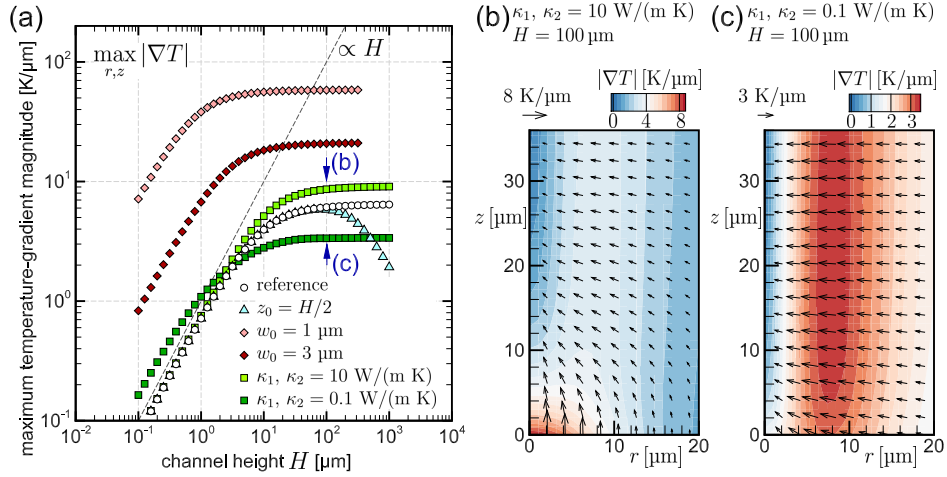}
    \caption{(a) Maximum of the magnitude of temperature gradient for various physical parameters, where unmentioned parameters are set to case A in Table~\ref{tab:phys-param}. (b,c) Temperature-gradient vectors and the heatmaps of their magnitude for (b) $\k1=\k2=10$~W/(m K) and (c) $\k1=\k2=0.1$~W/(m K) in panel (a) at $H=100$~\textmu m.}
    \label{fig:convection-gradT}
\end{figure*}
\begin{figure*}[bt]
    \centering
    \includegraphics[width=0.7\linewidth]{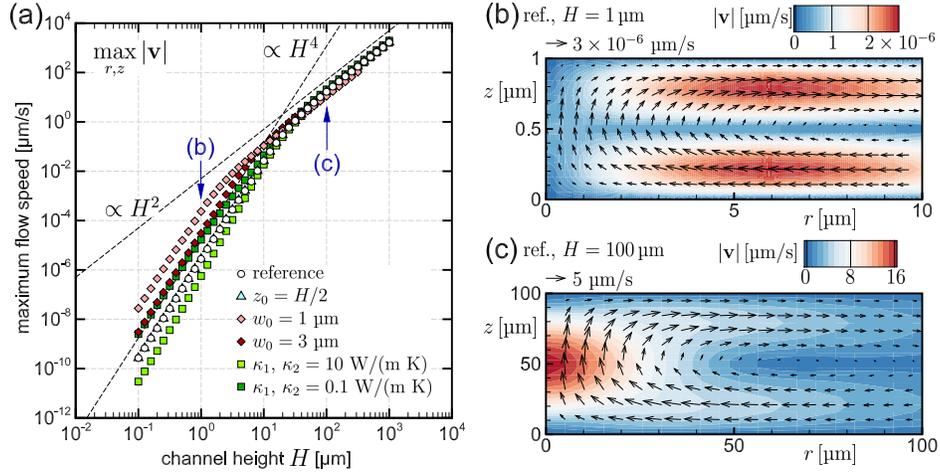}
    \caption{(a) Maximum flow speed for various physical parameters, where unmentioned parameters are set to case A in Table~\ref{tab:phys-param}. (b,c) flow velocity vectors and the heatmaps of their magnitude for (b) $H=1$~\textmu m and (c) $H=100$~\textmu m for the reference case in panel (a).}
    \label{fig:convection-v}
\end{figure*}

We then focus on the magnitude of the temperature gradient for the same parameter set. Figure~\ref{fig:convection-gradT}(a) shows the magnitude of the maximum temperature gradient $\max |\nabla T|$, where the legend is the same as that of Fig.~\ref{fig:convection-T}. 
First, the effect of $w_0$ shows that the magnitude of the temperature gradient increases in proportion to $H$ for $H<w_0$ and approaches constant values as $H$ becomes large. The positions at which the maximum occurs are on the beam axis $r=0$~\textmu m and on the surface of the solid $z=0$~\textmu m, except for the case of $\k1=\k2=0.1$~W/(m~K). To see this, Figs.~\ref{fig:convection-gradT}(b) and \ref{fig:convection-gradT}(c) show the temperature-gradient vector and its magnitude for the cases of (b) $\k1=\k2=10$ and (c) $\k1=\k2=0.1$~W/(m~K), respectively. Figure~\ref{fig:convection-gradT}(b) indicates that $|\nabla T|$ becomes maximum at the origin and is directed in the $z$ direction while $|\nabla T|$ becomes maximum near $r\approx 10$~\textmu m and $z\approx 25$~\textmu m in Fig.~\ref{fig:convection-gradT}(c), being directed to the $r$ direction. These features are consistent with those of the temperature field shown in Fig.~\ref{fig:convection-T}(f,\,g). Note that the temperature-gradient profile is important when thermal forces acting on dispersed objects (i.e., thermophoresis) are considered. As for the case of $z_0=H/2$ in Fig.~\ref{fig:convection-gradT}(a), i.e., the focal plane placed at the middle of the channel, $\max|\nabla T|$ decreases at large $H$. This is because the distance between the positions of the maximum temperature gradient (i.e., $r=z=0$~\textmu m) and the maximum temperature rise (i.e., the focal plane $z=H/2$) becomes larger as $H$. 

Finally, we show in Fig.~\ref{fig:convection-v}(a) the maximum flow speed, $\max|\v|$, for the same parameter set as in Figs.~\ref{fig:convection-T} and \ref{fig:convection-gradT}. 
Note that the flow magnitude less than $0.1$~\textmu m/s is so small that it may be practically undetectable in experiments. Therefore, the plots for roughly $H<10$~\textmu m in Fig.~\ref{fig:convection-v}(a) are physically irrelevant for the condition in Table~\ref{tab:phys-param}, but here we include these plots in the discussion below to investigate the fundamental flow characteristics. 

It is seen from Fig.~\ref{fig:convection-v}(a) that the flow speed increases in proportion to $H^4$ and $H^2$ for small $H$ and large $H$, respectively, where the transition between these two behaviors occurs near $H=w_0$ [see the plots of reference ($w_0=10$~\textmu m), $w_0=1$~\textmu m, and $w_0=3$~\textmu m]. 
To investigate the difference of these two trends, Fig.~\ref{fig:convection-v}(b,\,c) shows the corresponding flow velocity vectors and the heatmaps of its magnitude at (b) $H=1$~\textmu m and (c) $H=100$~\textmu m for the reference case in Fig.~\ref{fig:convection-v}(a). 
In addition to the almost $10^6$-order-of-magnitude difference between Figs.~\ref{fig:convection-v}(b) and \ref{fig:convection-v}(c), the positions of the maximum speed are different. That is, the maximum speed occurs away from the beam axis ($r=0$) for small $H$ while it is on the axis for large $H$. The latter is intuitively natural because the driving force for thermal convection is a buoyancy force in the $z$ direction, whereas the former is due to the tight confinement which suppresses the flow in the $z$ direction. 

\red{Unfortunately, it is difficult to explain explicitly the scaling transition from $|\u|\sim H^4$ to $|\u|\sim H^2$ using the semi-analytical solution, because naive power-series expansions of the semi-analytical solution with respect to $\tH(=H/w_0)$ do not work: $\tH$ appears in the semi-analytical solution together with $s$ [e.g., $\exp(s\tH)$], where $s$ is an integration variable in the inverse Hankel transform ranging from $0$ to $\infty$. We note that the origin of the scaling transition in a similar situation was explained in Sec.~IV.~B of Ref.~\cite{Riviere2016}. In Ref.~\cite{Riviere2016}, for small $H$ [i.e., the case with $\max|u_r|>\max|u_z|$; see Fig.~\ref{fig:convection-v}(b)], the scaling $u_r\sim H^3$ was derived using a lubrication approximation under the condition that the temperature increase does not depend on $H$. Since the temperature increase in our case scales as $\sim H$ as shown in Fig.~\ref{fig:convection-T}(a), we expect $|\u|\sim H^4$ for small $H$. On the other hand, for large $H$ [i.e., the case with $\max|u_z|>\max|u_r|$; see Fig.~\ref{fig:convection-v}(c)], Ref.~\cite{Riviere2016} derived the scaling $u_z\sim H^2$ using the balance between the viscous term and the buoyancy-force term. 
}

As for the case of $z_0=H/2$ in Fig.~\ref{fig:convection-v}(a), i.e., the focal plane placed at the middle of the channel, there is no difference compared with the reference. The cases of large (or small) thermal conductivities of the solids in Fig.~\ref{fig:convection-v}(a) results in smaller (or larger) flow speed for small $H$. However, for large $H$, the effect of thermal conductivity of the solid is less significant on the flow speed.


\subsubsection{Thermo-osmotic slip flow induced by a quasi point-like heat source}\label{sec:slip}
Using the thin-film approximation, we can imitate a point-like heat source such as Au plasmonic nanoparticle (say, a diameter of $\approx 100$~nm) attached to the channel wall, by reducing the beam width. In this section, we investigate the thermo-osmotic slip induced by this quasi point-like heat source. 

\begin{figure*}[bt]
    \centering
    \includegraphics[width=0.8\linewidth]{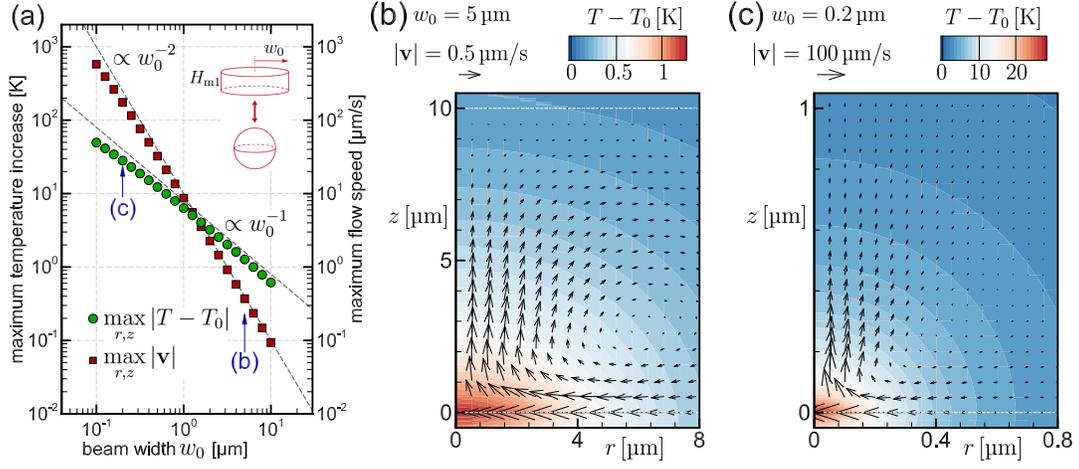}
    \caption{(a) Effect of beam width $w_0$ on the maximum temperature increase and the maximum flow speed for $H=10$~\textmu m and $\Hm1=0.05$~\textmu m, where unmentioned parameters are set to case B in Table~\ref{tab:phys-param}. (b,c) flow velocity vectors and the heatmaps of temperature increase for (b) $w_0=5$~\textmu m and (c) $w_0=0.2$~\textmu m in panel (a).}
    \label{fig:slip}
\end{figure*}

To do this modeling, we assume that the thin film 1 has a thickness of $\Hm1=0.05$~\textmu m and thermal conductivity $\km1=\k1(=1)$~W/(m K). Other parameters are set to those in case B in Table~\ref{tab:phys-param} unless otherwise stated. Suppose that we set the beam width to $w_0=0.2$~\textmu m. [Then, the ratio $\Hm1/w_0=0.25$, which is a small parameter $\ep$ in the thin-film approximation, results in the error of $\ep^2\approx 6$~\% according to the asymptotic analysis in \SI~A; this trend has been confirmed in Fig.~\ref{fig:Hm1_dependence}(b).] With this condition, the heat source, which is cylinderical in the thin film, has the approximate volume of $\pi w_0^2 \Hm1 \approx 6\times10^{6}$~nm$^3$. This means that the heat source has an equivalent radius of $r_{\mathrm{NP}}\approx 110$~nm if it is considered as a sphere [see the inset of Fig.~\ref{fig:slip}(a)]. Since the thermal conductivity of the thin film is matched to that of the solid 1 $(\km1=\k1)$, we can consider that the heat source of a nanoscale sphere with radius $r_{\mathrm{NP}}$ is embedded in the surface of the solid 1. From the channel-scale viewpoint with the order of micrometers, this spherical heat source can be regarded as a quasi point-like heat source. 

\begin{figure*}[bt]
    \centering
    \includegraphics[width=0.9\textwidth]{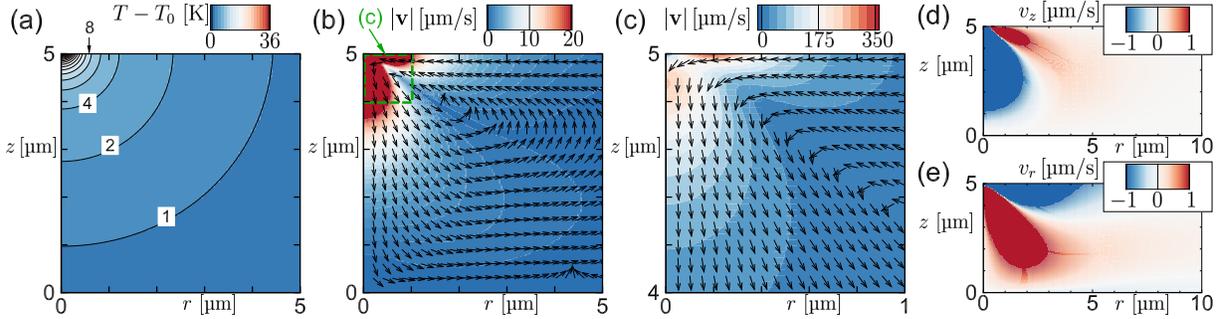}
    \caption{A demonstration of temperature and flow velocity fields for a quasi point-like heat source: a trial to reproduce Figs.~1(c), 2(b) in the main text of Ref.~\cite{Bregulla2016} and Fig.~6 in the supporting information of Ref.~\cite{Bregulla2016}. (a) Temperature increase; contours are separated by 2~K unless otherwise indicated (see Fig.~2(b) in Ref.~\cite{Bregulla2016} for comparison). Maximum value is $T-T_0\approx 26$~K. (b) Flow velocity field; vectors show only the direction of the flow and heatmaps indicate the magnitude (see Fig.~1(c) in Ref.~\cite{Bregulla2016} for comparison). Maximum value is $|\v|\approx 350$~\textmu m/s. (c) The magnification of the panel (b). (d,e) Heatmaps of (d) $v_z$ and (e) $v_r$; contour values are restricted to the range from $-1$ to $1$ to compare with Fig. 6 in the supporting information of Ref.~\cite{Bregulla2016}.}
    \label{fig:PRL2016}
\end{figure*}

Figure~\ref{fig:slip}(a) shows the maximum temperature increase (green circle) and the maximum flow speed (red square) for various beam width $w_0$. 
As described above, when $w_0$ becomes small, being comparable to the film thickness $\Hm1$, we can consider the heat source as point-like. As $w_0$ is decreased, the maximum values of the temperature increase and the flow speed are enhanced in proportion to $\propto w_0$ and $\propto w_0^2$, respectively. Figures~\ref{fig:slip}(b) and \ref{fig:slip}(c) display the corresponding temperature and flow fields for (b) $w_0=5$~\textmu m and (c) $w_0=0.2$~\textmu m in Fig.~\ref{fig:slip}(a). Note the difference of the displayed $rz$ ranges between Figs.~\ref{fig:slip}(b) and Fig.~\ref{fig:slip}(c). 
For smaller beam width $w_0$, it is seen that the temperature increase of $O(10)$~K leads to the flow speed of $O(100)$~\textmu m/s, and both the temperature and the flow are localized near the heat source. 

Let us now compare the present method using the quasi point-like heat source with the existing study that employed an Au nanoparticle as a heat source. 
In Ref.~\cite{Bregulla2016}, an Au nanoparticle with a radius of $r_{\mathrm{NP}}=125$~nm is attached to the upper surface of a micro channel ($H=5$~\textmu m) with the thickness of the solid parts $H_1=H_2=75$~\textmu m. The nanoparticle is heated by the laser and its temperature increase was estimated as $80$~K. 
The resulting temperature and flow fields was investigated in Ref.~\cite{Bregulla2016} and we use the value of the slip coefficient $\Kd2=-4.36$~\textmu m$^2$/(s K) reported there. 
(Note that $\chi$ in Ref.~\cite{Bregulla2016} is related to $\Kd2$ as $\Kd2=-\chi/T$.)
For other parameters, we use the values of $\kappa=0.6$~W/(m K), $\Hm1=0$~nm (i.e., no heat source at the bottom surface), $\Hm2=50$~nm, $\km2=1$~W/(m K), $z_0=H$, $w_0=0.3$~\textmu m, which is equivalent to $r_{\mathrm{NP}}\approx 150$~nm, $\Ab2=2\times10^4$~m$^{-1}$, and the rest of the parameters are set to those in case B of Table.~\ref{tab:phys-param}. 

\begin{figure*}[bt]
    \centering
    \includegraphics[width=0.8\textwidth]{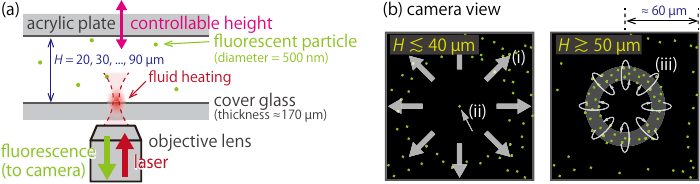}
    \caption{Schematic of the experiment in Ref.~\cite{Tsuji2021}. (a) A microslit containing fluorescent particles with a diameter of $500$~nm was irradiated by a near infrared laser from below. Fluorescence was observed from the bottom and recorded by a camera. The slit width $H$ varied from $H=20$~\textmu m to $90$~\textmu m. (b) Camera view for $H\leq 40$~\textmu m (left) and $H\geq 50$~\textmu m (right). The laser is irradiated at the center in the direction perpendicular to the view. }
    \label{fig:electro}
\end{figure*}

Figure~\ref{fig:PRL2016}(a) and ~\ref{fig:PRL2016}(b) show the temperature contours and flow velocity vectors, respectively, and Fig.~\ref{fig:PRL2016}(c) is the magnification of Fig.~\ref{fig:PRL2016}(b). Here, the vectors only display the flow direction, and the flow magnitude is described by the heatmap.  
Although the magnitudes of the temperature and flow velocity are slightly different from those in Ref.~\cite{Bregulla2016} (e.g., the maximum speed here is slightly overestimated), the shapes of the contours and vectors are in reasonable agreement (see Fig.~1(c) and 2(b) in Ref.~\cite{Bregulla2016}). 
It should be noted that $T-T_0$ is proportional to the inverse of the distance from the heat source: the Green's function for a point-heat source is $\sim \check{r}^{-1}$ with $\check{r}$ the distance from the heat source (see e.g., \cite{Govorov2007,Baffou2017}). To be more specific, Fig.~\ref{fig:PRL2016}(a) realizes this trend by observing, e.g., the contours for $T-T_0=1$, $2$, and $4$~K. Figures~\ref{fig:PRL2016}(d) and ~\ref{fig:PRL2016}(e) are the corresponding heatmaps of $v_z$ and $v_r$ that should be compared with Fig.~6 in the supporting information of Ref.~\cite{Bregulla2016}. The shapes of the heatmap, especially the borders of $v_z\gtrless0$ and $v_r\gtrless0$ are well reproduced by using the present simplified model of the point-like heat source. 

\red{It should be remarked that the effect of thermal convection in Fig.~\ref{fig:PRL2016} is negligible. In fact, recalling the decomposition \eqref{eq:Stokes-decompose}, we notice that the terms $\dTb{2}\duc\ubc{2}$ and $\dTb{2}\dud{2}\ubd{2}{2}$ represent thermal convection [superscript $(\c)$] and thermo-osmotic slip [superscript $(\dd2)$], respectively, both of which are induced by the heating of surface 2 [superscript $(\b2)$]. In the semi-analytical solution, we can separately evaluate these terms and it is confirmed that 
\begin{align*}
\frac{\max|\dTb{2}\duc\ubc{2}|}{\max|\dTb{2}\dud{2}\ubd{2}{2}|}\approx 3\times10^{-6}, 
\end{align*}
which indicates the effect of thermal convection is negligibly small compared with that of thermo-osmotic slip for the parameter set in Fig.~\ref{fig:PRL2016}. 
}

The limitation of the quasi point-like heat source should be described. If a heat source has nanoscale structures, such as bowtie nanoantenna \cite{Jones2018} or low-thermal conductivity plasmonic nanostructure \cite{Tamura2022a}, a temperature field cannot be considered as point-like near these structures. For such cases, a Green function approach \cite{Baffou2010a} is useful to obtain a temperature field semi-analytically, although a flow field requires some efforts on simulation. 
On the other hand, if a heat source is point-like and one is interested in larger scale of temperature and flow fields, the present method may be applicable even when the heat source is not necessarily originated from photothermal effects. For instance, a point-like heat may be created by the Joule heating of fabricated micro electrode pattern \cite{Chen2016}.

\subsubsection{Optothermal trap}\label{sec:optothermal-trap}

As the last example of the application of the present semi-analytical solution, we investigate the optothermal trapping of nanoparticles in a microslit \cite{Tsuji2021}. 
First, we explain the overview of the experimental results in Ref.~\cite{Tsuji2021}.
Figure~\ref{fig:electro}(a) shows the schematic of the experiment: a microslit with the slit width (i.e., channel height) $H$ contains a water solution with dispersed fluorescent polystyrene particles with a diameter of $500$~nm. The slit width $H$ was variable from $20$~\textmu m to $90$~\textmu m by a mechanical stage attached to the top acrylic plate. A near-infrared laser is loosely focused at the bottom of the slit using an objective lens, generating a heat in the fluid. The fluorescence of the particles are recorded by a camera through the same objective lens. Figure~\ref{fig:electro}(b) shows the resulting quasi-steady particle distribution at a fixed $H$: left (or right) shows the cases of $H\leq 40$~\textmu m (or $H\geq 50$~\textmu m), where the laser position is the center of the view. 
There was a clear threshold between $H=40$ and $50$~\textmu m in the particle behavior. That is, for $H\leq 40$~\textmu m, (i) most particles are depleted from the laser (i.e., the heated region), whereas (ii) a few particles were trapped at the very center of the view; for $H\geq 50$~\textmu m, (iii) the particles are trapped in a ring-like region (or possibly a toroid), showing rotating motion as shown in the figure. Moreover, the radius of the ring-like region increased as $H$. Hereafter, these three features are investigated using the semi-analytical solution.

\begin{figure*}[bt]
    \centering
    \includegraphics[width=0.8\textwidth]{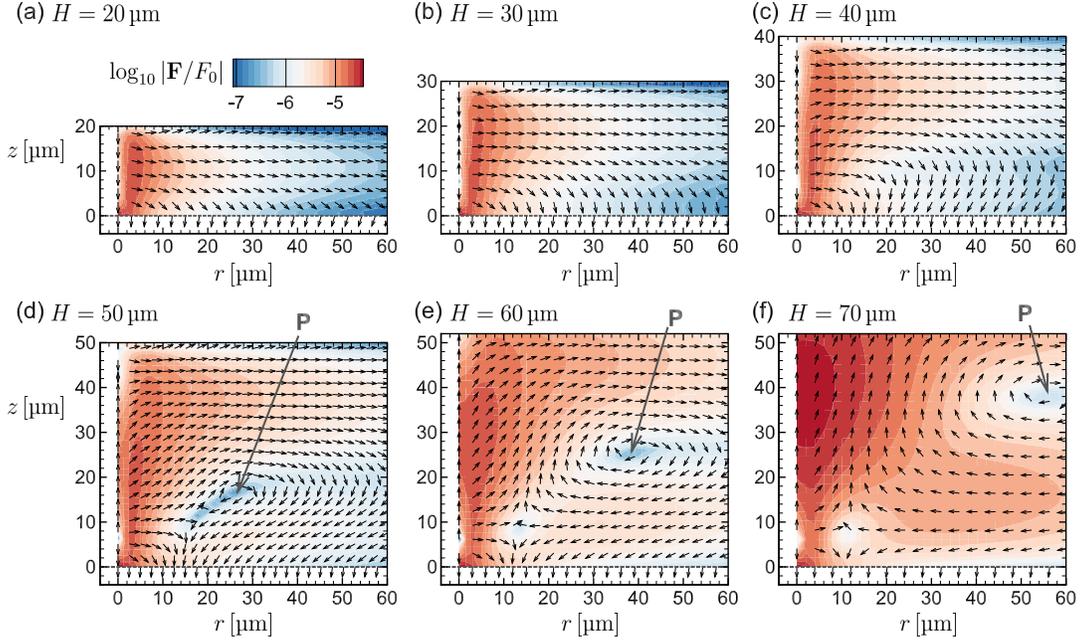}
    \caption{Force field acting on the particles using a physical parameter set in optothermal-trapping experiments \cite{Tsuji2021} with various height $H$ from (a) 20~\textmu m to (f) 70~\textmu m. Vector and heatmap show $\F/|\F|$ and $\log_{10}|\F/F_0|$, respectively, where $F_0$ is a reference magnitude of the force. Note that the vector length is uniform. The value $\log_{10}|\F/F_0|=-5$ indicates that $|\F|$ is $O(1)$ fN.}
    \label{fig:optothermal-logF}
\end{figure*}

\begin{figure*}[bt]
    \centering
    \includegraphics[width=0.8\textwidth]{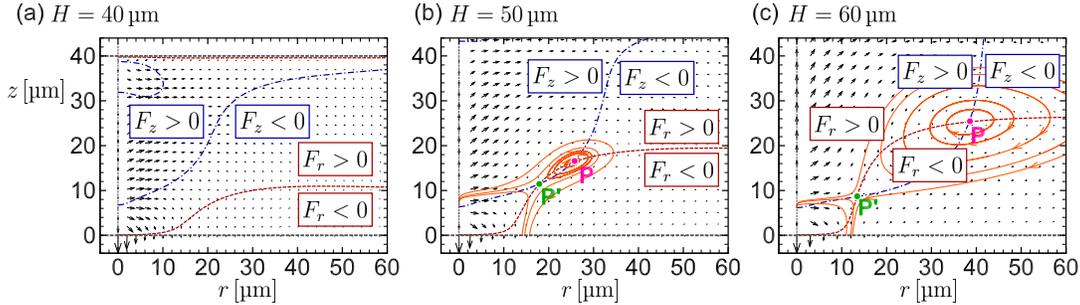}
    \caption{Drastic change of the force field between $H=40$~\textmu m and $H=50$~\textmu m. Force vector $\F=(F_r,\,F_z)$, the isolines of $F_r=0$ (red dashed) and $F_z=0$ (blue dash-dot), and the streamlines of $\F$ near P and P', at which the forces vanish, are presented for the cases of (a) $H=40$~\textmu m, (b) $H=50$~\textmu m, and (c) $H=60$~\textmu m. }
    \label{fig:optothermal-FrFz}
\end{figure*}

\begin{figure*}[bt]
    \centering
    \includegraphics[width=0.8\textwidth]{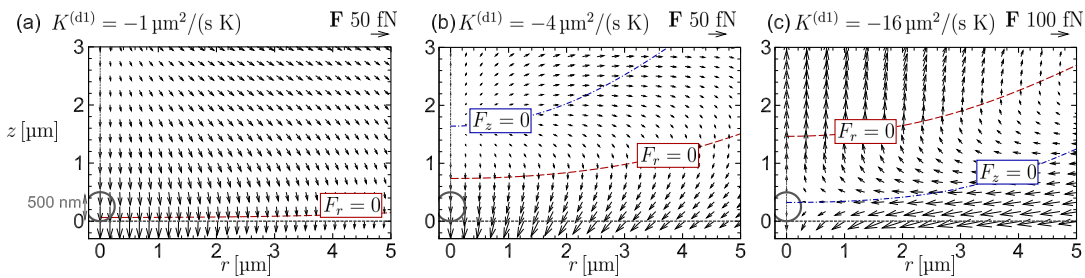}
    \caption{Effect of thermal slip coefficient $\Kd1$ on the optothermal trapping. Force vector $\F=(F_r,\,F_z)$, the isolines of $F_r=0$ (red dashed) and $F_z=0$ (blue dash-dot)  are presented for the cases of $H=40$~\textmu m and (a) $\Kd1=-1$, (b) $-4$, and (c) $-16$~\textmu m$^2$/(s K).  }
    \label{fig:optothermal-mag}
\end{figure*}

\begin{table*}[tb]
\centering
\def\valkap {$0.6$}  
\def\valkapa{$1.4$}
\def\valkapb{$0.15$}
\def\valH   {$20$--$90$}
\def\valHa  {$170$}
\def\valHb  {$10^3$}
\def\valw   {$4$}
\def\valP   {$4.6\times10^{-2}$}
\def\valDT   {$0.85$}
\def\valKda{$-1$ (or $-4$, $-16$)}
\def\valKdb{$-1$ (or $-4$, $-16$)}
\caption{Physical parameters used in Sec.~\ref{sec:optothermal-trap}. 
Other parameters are the same as those of case A in Table~\ref{tab:phys-param}.}
\label{tab:phys-param-electro}
{\tabcolsep = 0.5em
\small
\begin{tabular}{llllllll}
\hline 
$\kappa$~[W/(m\;K)]      &\valkap & &$H$~[\um]                & \valH  & &$P$~[W]                  &\valP   \\
$\kappa_1$~[W/(m\;K)]    &\valkapa& &$H_1$~[\um]              & \valHa & &$\Kd1$~[\um$^2$/(s\;K)]  &\valKda \\
$\kappa_2$~[W/(m\;K)]    &\valkapb& &$H_2$~[\um]              & \valHb & &$\Kd2$~[\um$^2$/(s\;K)]  &\valKdb \\
$w_0$~[\um]              &\valw   & &$D_T$~[\um$^2$/(s\;K)] & \valDT &                           &        \\
\hline
\end{tabular}
}
\end{table*}

In the following, we show the (approximate) force field Eq.~\eqref{eq:force} acting on the particles, which is the sum of the drag and the thermal force. 
Note that other forces caused by particle-wall interaction and/or inter-particle interaction may be significant in experiments but they are neglected here. On the other hand, the optical force is considered negligible, since the laser is loosely focused. The present discussion is limited to the contribution of the flow and temperature fields. 
It should be remarked the particles are so small that the acceleration can be neglected. Therefore, the observation of the force field is directly related to the particle motion without the Brownian motion. 

The physical parameters are set according to the experiment as summarized in Table~\ref{tab:phys-param-electro}. It should be noted that the parameter $D_T$ and $\Kdj$ are difficult to measure; we determine them empirically as follows. 
In Ref.~\cite{Tsuji2021}, a surfactant was added to the water solution. Therefore, the particle's surface and the channel's surface are both considered to be covered with the surfactant molecules, yielding similar slip coefficients. In Ref.~\cite{Tsuji2023}, it was shown that when the particle's surface has a thermal slip coefficient $K^{(\mathrm{p})}$, the thermophoretic mobility $D_T$ is given as $-(2/3)K^{(\mathrm{p})}\xi_s$ with $\xi_s=3\kappa/(2\kappa + \kappa_{\mathrm{p}})$ and $\kappa_{\mathrm{p}}$ the thermal conductivity of the particle. Here we put $\kappa_{\mathrm{p}}=0.2$~W/(m K) as the thermal conductivity of polystyrene and thus $\xi_s\approx 1.286$, leading to $D_T\approx0.85$~\textmu m$^2$/(s K) when $K^{(\mathrm{p})} = -1$~\textmu m$^2$/(s K). The value of $\Kdj(=K^{(\mathrm{p})})$ is determined so that the order of magnitude is the same as the thermal slip coefficient of the other polymer-coated surface \cite{Bregulla2016}.

Figure~\ref{fig:optothermal-logF} shows the direction of the force vector $\F$ [Eq.~\eqref{eq:force}] with the heatmap in the log-scale magnitude, $\log_{10}|\F/F_0|$, where $F_0 = 3\pi d \eta v_0(\approx 1.2$~nN$)$ is a reference magnitude of the force. 
The channel height $H$ is varied from (a) $20$~\textmu m to (f) $70$~\textmu m. 
For smaller height $H\leq 40$~\textmu m [Figs.~\ref{fig:optothermal-logF}(a--c)], the particles are subject to the forces in the positive $r$ direction, except near the bottom surface $z=0$~\textmu m. This is the case in which the thermophoretic force beats the fluid force, explaining the feature (i) in Fig.~\ref{fig:electro}(b). 
As the height $H$ increases [Figs.~\ref{fig:optothermal-logF}(d--f)], there arises a region in which the force almost vanishes, i.e., $\log_{10}|\F/F_0|\approx-7$, indicated by ``P" in the figure. 
The distance of P (say, the force-free region) from the laser increases as $H$. 

To understand the origin of the force-free region, Fig.~\ref{fig:optothermal-FrFz} shows the force vector $\F=(F_r,F_z)$ with the isolines of $F_r=0$ (red dashed curve) and $F_z=0$ (blue dash-dot curve) for (a) $H=40$, (b) $50$, and (c) $60$~\textmu m. It is seen that the curves for $F_r=0$ and $F_z=0$ do not intersect when $H=40$~\textmu m. However, as $H$ increases to $H=50$ or $60$~\textmu m, these two curves approach and start to intersect at two points P and P', generating the force-free points. 
It turns out that P' is a saddle point but P is the center of a vortex. In fact, Figs.~\ref{fig:optothermal-FrFz}(b,\,c) show the streamlines (orange curves) near P and P'. 
Therefore, the particles near P are expected to keep rotating around the force-free region. 
In the experiments, we observe from the bottom, and thus the particles are distributed in the ring-like (or a toroidal) region. This explains the feature (iii) in Fig.~\ref{fig:electro}(b) for $H\geq 50$~\textmu m. 

Finally, to investigate the feature (ii) in Fig.~\ref{fig:electro}(b), Fig.~\ref{fig:optothermal-mag} shows the magnification near $r=z=0$~\textmu m for $H=40$~\textmu m. 
Here, vectors indicate $\F$ and the isolines are $F_r=0$ (red dash) and $F_z=0$ (blue dash-dot); Figs.~\ref{fig:optothermal-mag}(a,b,c) show the case of (a) $\Kd1=-1$, (b) $-4$, and (c) $-16$~\textmu m$^2$/(s K), respectively. To trap the particle of diameter $500$~nm, which is schematically inserted in Fig.~\ref{fig:optothermal-mag}, $F_r<0$ and $F_z<0$ are necessary near the origin. It is seen that the case of $\Kd1=-1$~\textmu m$^2$/(s K) and $\Kd1=-16$~\textmu m$^2$/(s K) show the smaller region of $F_r<0$ and $F_z<0$, compared with the case of $\Kd1=-4$~\textmu m$^2$/(s K). This indicates that to achieve the trapping at the origin the moderate value of the slip coefficient $\Kd1$ (or its ratio to the thermophoretic mobility $D_T$) is supposed to be important.

Note that, without the slip, there never arises the region of $F_r<0$ unless the thermophoretic mobility $D_T$ is negative (i.e., negative thermophoresis). In the experiment \cite{Tsuji2021}, however, $D_T$ is obviously positive [the left schematic of Fig.~\ref{fig:electro}(b)], and thus we consider that the slip does the job in the optothermal trapping. Further analysis including collective particle dynamics together with systematic experiments will be future work. 

\red{It should be noted that when particles are very close to wall surfaces, as in the case of optothermal trap, some additional hydrodynamic effects manifest. First, a correction on the Stokes drag and a torque acting on the particles appear \cite{ONeill1964,Goldman1967a} (see also a text book \cite{Happel2012}, Chap. 7 ``Wall effect on the motion of a single particle"); these effects depend on the ratio $d/h$, where $d$ and $h$ are the particle diameter and the distance from the wall, respectively, and the effective drag is increased in such situation. In addition, there usually arise shear flows near walls; the force and torque acting on the particles occur due to the shear \cite{Goldman1967,Happel2012}. Remark that nonlinear and/or inertia effects are absent in this study, and thus lift forces arising from Magnus effect \cite{Rubinow1961} and Segre-Silberberg effect \cite{Segre1961} are neglected. 
}

\red{
Our analysis here do not take into account above-mentioned near-wall effects and should be considered qualitative when the particles are close to the walls. Nonetheless, we believe that the present form Eq.~\eqref{eq:force} is simple and enough useful to predict the particle motion qualitatively. 
}


\section{\label{sec:conclusion}Conclusion}
In this paper, we have provided semi-analytical temperature and flow fields induced by photothermal effects of fluids and/or boundary surfaces under the irradiation of a single focused Gaussian beam. Although we have neglected the nonlinear effects, which are supposed to be minor in micro- and nanofluidic conditions, the semi-analytical solution can give an instant answer to the temperature and flow characteristics, leading to the deeper understanding of the photothermal experiments and the search for more desirable experimental conditions. 
Note that the case of the irradiation of multiple laser is also possible to analyze by superposition, as long as the linearization assumption is not violated. 
The semi-analytical solution are provided as an open source \cite{Tsuji2024a} available to researchers without computational-fluid-dynamics tools.

The next step is to implement the motion of dispersed particles and molecules using the Brownian dynamics or Fokker-Planck type equations to understand optothermal manipulation, including time-dependent cases \cite{Kavokine2020}. Including an air-liquid interface (e.g., bubbles \cite{Setoura2017}) or a liquid-liquid interface \cite{Caciagli2020} will be also a useful extension because these interfaces results in rich physics and applications.

\begin{acknowledgments}
This work was supported by JSPS KAKENHI grant No.~20H02067, 22K18770, and 22K03924, and also by JST PRESTO grant No. JPMJPR22O7. We thank Mr. Shota Suzuki for testing the python code developed in this paper. 
\end{acknowledgments}


%

%

\clearpage

\onecolumngrid

\setcounter{page}{1}

\appendix

\renewcommand{\appendixname}{Supporting Information}

\begin{center}
{\Large
Supporting Information on \\
``\titleB"}
\\[1em]
{\large
Tetsuro Tsuji$^\ast$, Shun Saito, and Satoshi Taguchi\\[0.5em]
Graduate School of Informatics, Kyoto University, Kyoto 606-8501, Japan\\[0.5em]
$^\ast$tsuji.tetsuro.7x@kyoto-u.ac.jp
}
\end{center}

\setcounter{figure}{0}
\renewcommand\thefigure{S\arabic{figure}}   
\setcounter{table}{0}
\renewcommand\thetable{S\arabic{table}} 


\section{Derivation of boundary conditions on a thin film}\label{sec:thin-film}
\subsection{Setting}
In this section, we focus on the interface between the fluid and the solid 1, and construct model boundary conditions that approximate the presence of a heat absorbing thin film i.e., the film 1. 
The thin film 2 can be approximated in the same manner. Using these boundary conditions is beneficial since there is no need to prepare a fine gird system resolving the structure of the films that are much thinner than characteristic length scales of heat and fluid flows. Notations here are same as the main text unless otherwise stated.

\begin{figure}[b]
    \centering
    \includegraphics[width=0.4\textwidth]{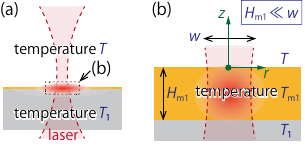}
    \caption{Schematic of the problem to derive the boundary condition (see Eq.~\eqref{eq:temperature-bc}) that approximates the presence of the thin film. (a) an entire view and (b) the magnification inside the thin film.}
    \label{fig:problem_thin-film}
\end{figure}

We consider a fluid-solid interface as shown in Fig.~\ref{fig:problem_thin-film}(a), where the thin film is attached on the solid. Magnifying Fig.~\ref{fig:problem_thin-film}(a) near the surface in the scale of the film thickness $\Hm1$, the situation can be described as Fig.~\ref{fig:problem_thin-film}(b), where we define $z=0$ at the interface between the fluid and the thin film 1. 
We assume that only the film absorbs the laser, since the case with an absorbing fluid can be superimposed. Then, the steady heat-conduction equation inside the film with a heat source due the laser is given as
\begin{align}\label{eq:film-eq}
\km1 (\Dp+\rnd{^2}{z^2}) \Tm1 + \Am1 I = 0 \quad ( -\Hm1< z < 0), 
\end{align}
where $\Tm1(r,z)$, $\km1$, and $\Am1$($=\Ab1$ in the main text) are the temperature, the thermal conductivity, and the absorption coefficient of the film, respectively, and $I(r,z)$ is the laser intensity distribution of a focused Gaussian beam defined in Eq.~\eqref{eq:laser} with a beam width $w(z)$; an operator $\Dp$ is defined in Eq.~\eqref{eq:Dp}.
Boundary conditions at two interfaces $z=0$ and $z=-\Hm1$ are 
\begin{subequations}\label{eq:film-bc}
\begin{align}
&\Tm1 = T_1, \quad \k1 \rnd{T_1}{z} = \km1 \rnd{\Tm1}{z} \quad (z=-\Hm1), \\
&\Tm1 = T, \quad \kappa \rnd{T}{z} = \km1 \rnd{\Tm1}{z} \quad (z=0), 
\end{align}
\end{subequations}
where $T$ and $T_1$ are the temperatures of the fluid $(z>0)$ and the solid $(z<-\Hm1)$, respectively; $\kappa$ and $\k1$ are the thermal conductivities of the fluid and solid, respectively. We assume that the film thickness $\Hm1$ is much smaller than the beam width $w(\geq w_0)$, namely, a parameter $\ep \equiv \Hm1/w_0$ is much smaller than unity, i.e., $\ep \ll 1$. In the following, we solve Eqs.~\eqref{eq:film-eq} and \eqref{eq:film-bc} and obtain the approximate solution $\Tm1$, deriving explicit relations between $T$ and $T_1$ and that between $\partial T/\partial z$ and $\partial T_1/\partial z$. These relations can be implemented as the boundary condition, where the effect of the thin-film heating is included. 

\subsection{Analysis}

Because of a tight confinement in the $z$ direction, the characteristic length scale of $\Tm1$ in the $z$ direction is $\Hm1$ and thus much smaller than that for the $r$ direction, i.e., the beam waist $w_0$. 
On the other hand, fluid and solid temperatures, $T$ and $T_1$, have slower variation of length scale, say, $w_0$. That is, we have the following scaling: 
\begin{align}
\rnd{\Tm1}{z} \sim O(\frac{\Tm1}{\Hm1}),  \quad \rnd{h}{z} \sim O(\frac{h}{w_0}) = \ep O(\frac{h}{\Hm1}) \quad (h=T,\,T_1). \label{eq:scaling}
\end{align}
Therefore, we introduce non-dimensional independent variables using $\Hm1$ as a reference length. To be more specific, we set $r = w_0 \tr$ and $z = -\hf\Hm1 + \Hm1 \tz$ (i.e., the interfaces are placed at $\tz=\pm\hf$). 
In the slower length scale, $z$ is expressed as $z=-\hf\Hm1 + w_0\tzs$, i.e., $\tzs = \ep \tz$.
In the following, we approximate Eqs.~\eqref{eq:film-eq} and \eqref{eq:film-bc} using the above scaling with $\ep=\Hm1/w_0\ll1$.

First, the beam width $w(z)$ (Eq.~\eqref{eq:beam-width}) is approximated. We consider that the distance between the focal plane $z=z_0$ and the film is the order of $w_0$ or smaller, because the laser intensity decreases and become insignificant as the distance increases further. Accordingly, $z_0$ is expressed as
$z_0= -\hf\Hm1 + w_0 \tz_0$ with a constant $|\tz_0|=O(1)$. Then, noting that the minimum beam waist $w_0$ is typically $w_0\geq \lambda/2$ in experiments, we can expand $w(z)$ with respect to $\ep\ll1$ as
\begin{subequations}
\begin{align}
&\frac{w(z)}{w_0} \equiv \tw(\tz) = \twm1 \left(1+ a\s1\ep \tz  + O(\ep^2)\right), \label{eq:wa} \\
&\twm1 = (1+C_w \tz_0^2)^{1/2}, \quad 
C_w\equiv \lambda^2/(\pi w_0)^2, \quad 
a\s1 = -\tz_0^{-1}(1-\twm1^{-2}),
\label{eq:wa-for-case1}
\end{align}
\end{subequations}
where $C_w = O(1)$ is a constant and $\twm1=\wm1/w_0=O(1)$ is the non-dimensional version of the beam radius $\wm1$ at the center of the film $z=-\hf\Hm1$. Therefore, we can consider that the beam has a constant radius $\wm1$ inside the film at the leading order. The beam intensity (Eq.~\eqref{eq:beam-intensity}) is readily expanded in the same manner as
\begin{subequations}
\begin{align}
&\frac{I}{I_0}  =  \tIm1(\tr)\left(1+b\s1(\tr) \ep \tz 
 + O(\ep^2)\right), \label{eq:Ib} \\
&
\tIm1 =\frac{1}{\twm1^{2}} \exp(-\frac{2\tr^2}{\twm1^2}), \quad 
b\s1 = 2 a\s1 \left( \frac{2}{\twm1^2}\tr^2-1\right).   
\label{eq:Ib-for-case1}
\end{align}
\end{subequations}

Next, we look for the approximate solution of $\Tm1$. To this end, we introduce non-dimensional quantities as
\begin{equation}
    \begin{split}
&\Tm1(r,z) = T_0\left(1 + \dm1 \tm1(\tr,\tz)\right)\quad \left(\dm1 = \frac{\Am1 I_0 w_0^2/\km1}{ T_0}\right), \\
&
T(r,z) = T_0\left( 1 + \dm1 \tau(\tr,\tzs)\right), \quad 
T_1(r,z) = T_0\left(1 + \dm1 \t1(\tr,\tzs)\right), \\
&\kappa = \alpha   \km1, \quad 
\k1    = \alpha_1 \km1, \quad 
\tDp   = \frac{1}{\tr}\rnd{}{\tr}(\tr\rnd{}{\tr}).
    \end{split}
    \label{eq:temperature-film-nd}
\end{equation}
Substituting Eq.~\eqref{eq:temperature-film-nd} into Eqs.~\eqref{eq:film-eq} and \eqref{eq:film-bc}, we obtain
\begin{subequations}\label{eq:film-nd}
\begin{align}
&0 =  (\ep^2\tDp+\rnd{^2}{\tz^2}) \tm1(\tr,\tz)  + \ep^2\tIm1(\tr)\left(1+b\s1 \ep \tz + \cdots\right) \quad \left( -\hf \leq \tz \leq \hf\right), \\
&\mathrm{b.c.}\quad \begin{cases}
\tm1 = \t1, \quad \alpha_1 \ep \drnd{\t1}{\tzs} = \drnd{\tm1}{\tz} \quad (\tz=-\dhf), \\[0.5em]
\tm1 = \tau, \quad \alpha \ep  \drnd{\tau}{\tzs} = \drnd{\tm1}{\tz} \quad (\tz=\dhf), 
\end{cases}
\end{align}
\end{subequations}
where the scaling in Eq.~\eqref{eq:scaling} is taken into account, i.e., $\partial \tm1/\partial \tz=O(1)$ but $\partial h/\partial \tz=\ep(\partial h/\partial \tzs) = O(\ep)$ $(h=\tau,\,\t1)$. 

Now, expanding $\tm1$, $\tau$, and $\t1$ in the power series of $\ep$ as 
$h = h\s0 + \ep h\s1 + \ep^2 h\s2 + \cdots$ $(h=\tm1$, $\tau$, $\t1)$, and substituting these into Eq.~\eqref{eq:film-nd}, 
at the $0$-th order in $\ep$ we obtain
\begin{subequations}
\begin{align}
&\rnd{^2\tm1\s0}{\tz^2} =0, \\
& \mathrm{b.c.}\quad \tm1\s0 = \t1\s0 \quad (\tz=-\hf) \text{\quad and \quad} \tm1\s0 = \tau\s0 \quad (\tz=\hf), 
\\
& 
\mathrm{b.c.}\quad 0 = \drnd{\tm1\s0}{\tzs} \quad (\tz=-\hf) 
\text{\quad and \quad}
0 = \drnd{\tm1\s0}{\tzs} \quad (\tz=\hf),  
\end{align}
\end{subequations}
which yields a trivial solution $\tm1\s0=\tau\s0=\t1\s0$. That is, we obtain a usual Dirichlet condition $T=T_1$ at the surface in the limit of the zero film thickness (i.e., $\ep=\Hm1/w_0\to0$).
On the other hand, the $\ell$-th order systems with $\ell=1,2,3$ are rearranged as 
\begin{subequations}\label{eq:film-nd-power}
\begin{align}
&
0 = \rnd{^2\tm1\sl}{\tz^2}  + \mI\sl(\tr,\tz) , \quad 
( -\hf\leq \tz \leq \hf), 
\label{eq:film-nd-power-eq}\\
&\text{b.c.}\quad 
\tm1\sl = \t1\sl \quad (\tz=-\hf) \text{\quad and \quad} \tm1\sl = \tau\sl \quad (\tz=\hf), 
\label{eq:film-nd-power-bc1}
\\
&\text{b.c.}\quad 
\alpha_1 \drnd{\t1\slm1}{\tzs} = \drnd{\tm1\sl}{\tz} \quad (\tz=-\hf) 
\text{\quad and \quad}
\alpha  \drnd{\tau\slm1}{\tzs} = \drnd{\tm1\sl}{\tz} \quad (\tz=\hf),  
\label{eq:film-nd-power-bc2}
\end{align}
\end{subequations}
where $\mI\sl$ $(\ell=1,\,2,\,3)$ are the inhomogeneous terms: 
\begin{align}
\mI\s1 = 0, \quad 
\mI\s2 = \tIm1(\tr) + \tDp\tm1\s0, \quad 
\mI\s3 = b\s1\tIm1(\tr) \tz + \tDp\tm1\s1. 
\end{align}
These systems for $\ell=1,\,2,\,3$ can be solved from the lowest order in $\ep$. 
In fact, Eqs.~\eqref{eq:film-nd-power-eq} and \eqref{eq:film-nd-power-bc1} can be solved explicitly as
\begin{align}
\tm1\sl(\tr,\tz) = \tau_\dif\sl(\tr) \tz + \tau_\ave\sl(\tr) - \mIs\sl (\tr,\tz)\quad (\ell=1,\,2,\,3), \label{eq:approx-sol}
\end{align}
where shorthand notations are introduced as
\begin{align}
\tau_\dif\sl(\tr) = \tau\sl\subp -\t1\sl\subm, \quad 
\tau_\ave\sl(\tr) = \hf(\tau\sl\subp  + \t1\sl\subm) \quad (\ell=0,...,3), 
\end{align}
(note that $\tau_\dif\s0 =0$) and $\mIs\sl$ $(\ell=1,\,2,\,3)$ are inhomogeneous solutions including the generated heat: 
\begin{subequations}\label{eq:inhomo-sol}
\begin{align}
 &\mIs\s1 = 0, \\
 &\mIs\s2 = \hf(\tz^2-\frac{1}{4}) (\tIm1(\tr) +\tDp \tau_\ave\s0), \\
 &\mIs\s3 = \frac{1}{6}\tz(\tz^2-\frac{1}{4})(b\s1\tIm1(\tr)+\tDp\tau_\dif\s1)
+ 
  \hf(\tz^2-\frac{1}{4})\tDp \tau_\ave\s1.
\end{align}
\end{subequations}
Substituting the approximated solution (Eq.~\eqref{eq:approx-sol}) into Eq.~\eqref{eq:film-nd-power-bc2}, we obtain constraints for the $\tz$ derivatives: 
\begin{align}
\alpha_1 \rnd{\t1\slm1}{\tzs}\bsubm = \tau_\dif\sl - \rnd{\mIs\sl}{\tz}\bsubm \text{\quad and \quad}
\alpha \rnd{\tau\slm1}{\tzs}\bsubp = \tau_\dif\sl - \rnd{\mIs\sl}{\tz}\bsubp, 
\end{align}
which results in the direct relations between the quantities of fluid and those of solid for $\ell=1,\,2,\,3$: 
\begin{align}
\begin{cases}
\tau_\dif\sl = \dhf \left( \alpha\drnd{\tau\slm1}{\tzs}\bsubp + \alpha_1 \drnd{\t1\slm1}{\tzs}\bsubm \right) 
+ \mA\sl, \quad \mA\sl \equiv 
\dhf\left( \rnd{\mIs\sl}{\tz}\bsubp+\rnd{\mIs\sl}{\tz}\bsubm \right), \\[1em]
\alpha\drnd{\tau\slm1}{\tzs}\bsubp - \alpha_1 \drnd{\t1\slm1}{\tzs}\bsubm = \mB\sl, \quad \mB\sl \equiv 
-\drnd{\mIs\sl}{\tz}\bsubp+\drnd{\mIs\sl}{\tz}\bsubm,
    \end{cases}
    \label{eq:film-result-ell}
\end{align}
where $\tm1$ is no more present. 
Recalling the expansion $h = h\s0 + \ep h\s1 + \ep^2 h\s2 + \cdots$ $(h=\tau$ and $\t1)$, we arrive at 
\begin{align}
\begin{cases}
\tau\subp-\t1\subm = \dhf \ep \left( \alpha\drnd{\tau}{\tzs}\bsubp + \alpha_1 \drnd{\t1}{\tzs}\bsubm \right) 
+ \ep \mA\s1
+ \ep^2 \mA\s2
 + \ep^3 \mA\s3 + O(\ep^4), \\[1em]
\alpha\drnd{\tau}{\tzs}\bsubp - \alpha_1 \drnd{\t1}{\tzs}\bsubm = \mB\s1 + \ep\mB\s2 + \ep^2\mB\s3 + O(\ep^3), 
    \end{cases}
    \label{eq:film-result-sum}
\end{align}
where $\mA\sl$ and $\mB\sl$ can be explicitly obtained as
\begin{subequations}\label{eq:AB-case1}
\begin{align}
&\mA\s1 = \mA\s2 = 0, \quad 
\mA\s3 = \frac{1}{12}(b\s1 \tIm1 + \tDp \tau_\dif\s1), \\
&
\mB\s1 = 0, \quad 
\mB\s2 = -\tIm1 - \tDp \tau_\ave\s0, \quad 
\mB\s3 = - \tDp \tau_\ave\s1. 
\end{align}
\end{subequations}
Note that Eq.~\eqref{eq:film-result-sum} recover usual boundary conditions at the interface of fluid and solid when $\ep\to0$ since $\mB\s1=0$. Neglecting the terms of $O(\ep^3)$, Eq.~\eqref{eq:film-result-sum} can be recast further as 
\begin{align}
\begin{cases}
\tau\subp-\t1\subm = \dhf \ep \left( \alpha\drnd{\tau}{\tzs}\bsubp + \alpha_1 \drnd{\t1}{\tzs}\bsubm \right) 
+ O(\ep^3), \\[1em]
\alpha\drnd{\tau}{\tzs}\bsubp - \alpha_1 \drnd{\t1}{\tzs}\bsubm = -\ep \left(\tIm1+\dhf\tDp(\tau\subp+\t1\subm)\right) + O(\ep^3), 
    \end{cases}
    \label{eq:film-result-sum2}
\end{align}
where the effect of the beam-width variation, i.e., $b^{(1)}$ in Eq.~\eqref{eq:Ib}, is $O(\ep^3)$ and thus can be neglected. 

\subsection{Summary}
\begin{figure}
    \centering
    \includegraphics[width=0.4\textwidth]{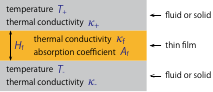}
    \caption{Schematic of the summary of \SI~A and Eq.~\eqref{eq:film-result-sum-dim2}. }
    \label{fig:problem_thin-film-general}
\end{figure}

In the dimensional form, the relation between the fluid temperature $T$ at the fluid-film interface $(z=0)$ and the solid 1 temperature $T_1$ at the solid-film interface $z=-\Hm1$ are related as 
\begin{align}
\begin{cases}
T\subpd-T_1\submd = 
\dhf\dfrac{\Hm1}{\km1} \left( \kappa\drnd{T}{z}\bsubpd + \k1 \drnd{T_1}{z}\bsubmd \right), \\[1em]
\kappa\drnd{T}{z}\bsubpd - \k1 \drnd{T_1}{z}\bsubmd = 
-\Am1\Hm1 I|_{z=-\hf\Hm1}  - \dhf\km1\Hm1\Dp(T+T_1).
    \end{cases}
    \label{eq:film-result-sum-dim}
\end{align}
Practically, there is no necessary to distinguish the upper surface $z=0$ and the lower surface $z=-\Hm1$ of the film when we are not interested in the temperature profile inside the film. 
Therefore, to generalize Eq.~\eqref{eq:film-result-sum-dim}, we introduce the $z$ position of the film as $\zf$, the thermal conductivity of the film as $\kf$, the thickness of the film as $\Hf$, the absorption coefficient of the film as $\Af$, the temperature and the thermal conductivity of a material touching to the top (or bottom) surface of the film as $T_+$ (or $T_-$) and $\kappa_+$ (or $\kappa_-$), respectively, as shown in Fig.~\ref{fig:problem_thin-film-general}. Then, Eq.~\eqref{eq:film-result-sum-dim} can be replaced by, neglecting terms of $O(\ep^3)$,  
\begin{align}
\begin{cases}
T_+-T_- = 
\dhf\dfrac{\Hf}{\kf} \left( \kappa_+\drnd{T_+}{z} + \kappa_- \drnd{T_-}{z} \right), \\[1em]
\kappa_+\drnd{T_+}{z} - \kappa_- \drnd{T_-}{z} = 
-\Af\Hf I(r,z)  - \dhf\kf \Hf\left(\dfrac{1}{r}\rnd{}{r}(r\rnd{}{r})\right)(T_++T_-), 
    \end{cases} (z=\zf).
    \label{eq:film-result-sum-dim2}
\end{align}
 
\section{Hankel transform and its application}\label{sec:hankel-application}
In the present axisymmetric problem, the Hankel transform is a useful tool to carry out the analysis. In the following, we first introduce the Hankel transform and then show its application to the present problem. 

For a function $f(r)$, the Hankel transform of the order $n$ and its inverse transform are defined as 
\begin{align}
\barf_n(s) = \int_0^\infty r f(r) J_n(rs) \dd r, \quad 
f(r) =  \int_0^\infty s \barf_n(s) J_n(rs) \dd s,
\end{align}
respectively, where $J_n$ is the Bessel function of the first kind. 
Some relations used in this paper are summarized in Table~\ref{tab:hankel}.
\begin{table}[tb]
\tabcolsep = 1em
    \centering
\renewcommand\arraystretch{2}
\caption{Some relations of Hankel transforms used in this paper. }
    \begin{tabular}{llll}
    \hline
index &order & original function & transformed function \\
\hline
1 &$0$   & $f(r)$ & $\barf_0(s)$ \\
2 &$0$   & $\drnd{f}{r}+\dfrac{f}{r}$ & $s \barf_1(s)^\dagger$ \\
3 & $0$   & $\drnd{^2f}{r^2}+\dfrac{1}{r}\drnd{f}{r}$& $-s^2 \barf_0(s)$ \\
4 &$0$   & $\dfrac{1}{2 a^2}\exp(-\dfrac{r^2}{4a^2})$& $\exp(-a^2s^2)$ \\
5 &$1$   & $\drnd{f}{r}$ & $-s \barf_0(s)$ \\
6 &$1$   & $\drnd{^2f}{r^2}+\dfrac{1}{r}\drnd{f}{r}-\dfrac{f}{r^2}$& $-s^2 \barf_1(s)$ \\[0.5em]
\hline
\multicolumn{4}{l}{
$^\dagger$ $\lim_{r\to\infty}(r f(r) J_0(sr))=0$ is assumed for any positive $s$. }
    \end{tabular}
    \label{tab:hankel}
\end{table}

To apply the Hankel transform to the present problem (Eqs.~\eqref{eq:temperature-nd} and \eqref{eq:Stokes-nd}), 
we define the Hankel transform of $\tx\,$, $\tx1$, $\tx2$, $\uxy$, and $\tpxy$ as 
\begin{subequations}
\begin{align}
&\bar{F}(s,\tz) = \int_0^\infty \tr F(\tr,\tz) J_0(s\tr) \dd \tr\quad 
(F=\tx\,,\tx1,\,\tx2,\,\uzxy,\,\tpxy), \\
&\burxy(s,\tz) = \int_0^\infty \tr \urxy(\tr,\tz) J_1(s\tr) \dd \tr.
\end{align}
\end{subequations}
The transformed functions (e.g., $\btx\,(s,\tz)$) can be obtained analytically as shown in Sec.~C. 
Once the transformed functions are obtained, the original functions can be recovered by the inverse transforms
\begin{subequations}
\begin{align}
&F(\tr,\tz) = \int_0^\infty s \bar{F}(s,\tz) J_0(s\tr) \dd s\quad 
(F=\tx\,,\tx1,\,\tx2,\,\uzxy,\,\tpxy), \\
&\urxy(\tr,\tz) = \int_0^\infty s \burxy(s,\tz) J_1(s\tr) \dd s.
\end{align}    
\end{subequations}

Now, applying the Hankel transform to Eqs.~\eqref{eq:temperature-nd} and \eqref{eq:Stokes-nd}, we obtain the set of ordinary differential equations for the temperature field as 
\begin{subequations}\label{eq:temperature-ht}
\begin{align}
&
-s^2 \btx\, + \rnd{^2\btx\,}{\tz^2}  + \bmIx = 0, \\
&-s^2 \btxj  + \rnd{^2\btxj}{\tz^2} = 0 \quad (j=1,\,2)\\
&
\begin{cases}
\btx\, - \btx1 = \dhf\dfrac{\tHm1}{\tkm1}
\left(\drnd{\btx\,}{\tz}+\tk1\drnd{\btx1}{\tz}\right), \\[1em]
\drnd{\btx\,}{\tz}-\tk1\drnd{\btx1}{\tz} = - \Big(\bmIx_1 - \dhf\tkm1 s^2 ( \btx\, + \btx1)\Big)\tHm1, 
\end{cases}
(\tz=0), \\[1em]  
&
\begin{cases}
\btx2 - \btx\, = \dhf\dfrac{\tHm2}{\tkm2}
\left(\tk2\drnd{\btx2}{\tz}+\drnd{\btx\,}{\tz}\right), \\[1em]
\tk2\drnd{\btx2}{\tz}-\drnd{\btx\,}{\tz} = - \Big( \bmIx_2 - \dhf\tkm2 s^2 (\btx2 + \btx\,)\Big)\tHm2, 
\end{cases}
(\tz=\tH),  \\[1em]
&\btx1 = 0 \quad (\tz=-\tH_1), \quad 
\btx2 = 0\quad (\tz=\tH+\tH_2), \\
&\bmIa = \bmIb1_1 = \bmIb2_2 = \frac{1}{4}\exp(-\frac{s^2\tw^2}{8}), \\
&\bmIa_1 = \bmIa_2=\bmIb1=\bmIb2=\bmIb1_2=\bmIb2_1=0.  
\end{align}
\end{subequations}
The conditions for the radial direction (Eq.~\eqref{eq:temperature-nd-bc-r}) are automatically satisfied, because the Bessel function vanishes as $J_1(0)=0$ and $\lim_{\tr\to\infty}J_1(s\tr)=0$ in Eq.~\eqref{eq:solution}.
For the flow field, we obtain the transformed equations as
\begin{subequations}\label{eq:Stokes-ht}
\begin{align}
&s\burxy+\rnd{\buzxy}{\tz}=0,\\
&
-s\btpxy =  -s^2\burxy + \rnd{^2 \burxy}{\tz^2}, \\
&
\rnd{\btpxy}{\tz} =  -s^2 \buzxy + \rnd{^2 \buzxy}{\tz^2} + \bmJxy, \\[0.5em]
&\begin{cases}
\burxy =  \bmJxy_1, \quad  \buzxy = 0 \quad (\tz=0), \\[0.5em]
\burxy =  \bmJxy_2, \quad  \buzxy = 0 \quad (\tz=\tH), 
\end{cases}\\
&\bmJxc = \btx\, , \quad 
\bmJxd1_1 = s\btx\,, \quad 
\bmJxd2_2 = s\btx\,, \\
&\bmJxd1 = \bmJxd2 = \bmJxc_1 = \bmJxd2_1 = \bmJxc_2 = \bmJxd1_2 = 0. 
\end{align}
\end{subequations}
In the same manner as the temperature, the conditions for the radial direction (Eq.~\eqref{eq:Stokes-nd-bc-r}) are automatically satisfied, because the Bessel function $J_0$ also vanishes at infinity in Eq.~\eqref{eq:solution}.
These are linear ordinary differential equations with inhomogeneous terms $\mIx$, $\mIx_1$, $\mIx_2$, $\mJxy$, $\mJxy_1$, and $\mJxy_2$ and can be solved explicitly as shown in the Supporting Information Sec.~C.

\section{Detail of the semi-analytical solutions} \label{sec:semi-anal-detail}
\subsection{Temperature field}
First, Eq.~\eqref{eq:temperature-ht} can be solved for $\x=\a$ (fluid heating) and $\x=\b j$ (surface $j$ heating) as 
\begin{subequations}\label{eq:temperature-solution}
\begin{align}
&\btx\, = \tausx \sum_{\alpha = \pm} \left(  \Bx{\alpha} +  \EEx(\alpha s, \tz -\tz_0) \right) e^{\alpha s(\tz-\tz_0)}, \label{eq:temperature-solution-fluid}\\
&\rnd{\btx\,}{\tz} = \tausx \sum_{\alpha = \pm} \left(  \Bx{\alpha} +  \EEx(\alpha s, \tz -\tz_0) - \GGx(\alpha s, \tz -\tz_0) \right) \alpha s e^{\alpha s(\tz-\tz_0)}, \\
&\btx1 =  \tausx\sum_{\alpha = \pm} \Bx{1}\alpha e^{\alpha s(\tz + \tH_1)}, \quad 
\btx2 = \tausx\sum_{\alpha = \pm}  \Bx{2}  \alpha e^{\alpha s(\tz - \tH-\tH_2)}, 
\end{align}
\end{subequations}
where $\sum_{\alpha=\pm}$ means the summation over $\alpha=+1$ and $\alpha=-1$; the coefficients $\Bx{\pm}$, $\Bx{1}$, $\Bx{2}$, and $\tausx$ are the functions of $s$; $\EEx$ and $\GGx$ are the inhomogeneous solutions expressed by explicit functions, as introduced below. 

The functions $\tausx$, $\EEx$, and $\GGx$ in Eq.~\eqref{eq:temperature-solution} are defined as
\begin{equation}
\begin{split}
&
\begin{cases}
\tausa(s) = \frac{\spi}{8 s^2} \exp(\tzR^2-\frac{s^2}{8}), \\
\EEa(s,\tz) = \tzR\,\erfc(\tzR+\frac{s\tz}{2\tzR}), \quad 
\GGa(s,\tz) = \frac{1}{\spi} \exp\left(-(\tzR+\frac{s\tz}{2\tzR})^2\right),
\end{cases}\\
&\tausb1(s) = \tfrac{\tHm1}{8s}\exp(-\tfrac{s^2 \tw^2(0)}{8}), \quad \EEb1=\GGb1 = 0, \\
&\tausb2(s) = \tfrac{\tHm2}{8s}\exp(-\tfrac{s^2 \tw^2(\tH)}{8}), \quad \EEb2=\GGb2 = 0, 
\label{eq:taus}
\end{split}
\end{equation}
where, in the case of fluid heating $(\x=\a)$, the inhomogeneous terms $\EEa$ and $\GGa$ appear. 

The integration coefficients in Eq.~\eqref{eq:temperature-solution} depend on $s$ and are given as
\begin{equation}
\begin{split}
&\Bx{\mp} 
= \QQ^{-1} ( \PP_{\mp,1}^{(\x)}\asfH \bsf0 \tk1
            +\PP_{\mp,2}^{(\x)}\asf0 \bsfH \tk2
            +\PP_{\mp,3}^{(\x)}\bsf0 \bsfH \tk1 \tk2 
            +\PP_{\mp,4}^{(\x)}\asf0 \asfH),
            \\
&\Bx{1} 
= \QQ^{-1} ( \PP_{1,1}^{(\x)} \bsfH \tk2
            +\PP_{1,2}^{(\x)} \asfH), \quad 
\Bx{2} 
= \QQ^{-1} ( \PP_{2,1}^{(\x)} \bsf0 \tk1
            +\PP_{2,2}^{(\x)} \asf0).
            \label{eq:B}
\end{split}
\end{equation}
Here, $\asf0$, $\asfH$, $\bsf0$, and $\bsfH$ are the functions of $s$ and depend on the thickness of solid $j$; In particular,
\begin{align}
\asf0 = 2\sinh s\tH_1, \quad 
\asfH = 2\sinh s\tH_2, \quad 
\bsf0 = 2\cosh s\tH_1, \quad 
\bsfH = 2\cosh s\tH_2.
\end{align}
The denominator $\QQ$ in Eq.~\eqref{eq:B} is defined as
\begin{align}
&\QQ = 
- e^{s\tH}
((\HH_{1\up}+\HH_{1+})\asf0+(\HH_{1\lw}+\HH_{1+})\bsf0\tk1)
((\HH_{2\up}+\HH_{2+})\asfH-(\HH_{2\lw}+\HH_{2+})\bsfH\tk2) \notag\\
&\phantom{Q = }+ e^{-s\tH}
((\HH_{1\up}-\HH_{1+})\asf0-(\HH_{1\lw}-\HH_{1+})\bsf0\tk1)
((\HH_{2\up}-\HH_{2+})\asfH+(\HH_{2\lw}-\HH_{2+})\bsfH\tk2), 
\end{align}
where $\HH_{j\pm}$, $\HH_{j \up}$, and $\HH_{j \lw}$ are the auxiliary notations related to the film parameters defined as
\begin{align}
&\HH_{j\pm} = 1\pm\frac{\tHmj^2}{4} s^2, \quad 
\HH_{j \up} = \tHmj \tkmj s, \quad 
\HH_{j \lw} = (\tHmj/\tkmj) s. \label{eq:H}
\end{align}
The coefficients $\PP_{\mp,p}^{(\x)}$ $(p=1,...,4)$ and $\PP_{q,p}^{(\x)}$ $(p,q=1,2)$ in Eq.~\eqref{eq:B} are lengthy and listed in Table~\ref{tab:P-tem}, in which the functions $\Iz$, $\IH$, $\dIz$, and $\dIH$ representing the inhomogeneous terms are related with the beam parameters $\tz_0$ and $\tzR$ through $\EEa$ and $\GGa$:
\begin{equation}
\begin{split}
&\Iz  = \sum_{\alpha=\pm}\EEa(\alpha s,-\tz_0) e^{-\alpha s \tz_0}, \quad 
\IH  = \sum_{\alpha=\pm}\EEa(\alpha s,\tH-\tz_0) e^{\alpha s(\tH- \tz_0)}, \\
&\dIz = \sum_{\alpha=\pm}\alpha\left(\EEa(\alpha s,-\tz_0)-\GGa(\alpha s, -\tz_0)\right) e^{-\alpha s\tz_0}, \\ 
&\dIH = \sum_{\alpha=\pm}\alpha\left(\EEa(\alpha s,\tH-\tz_0)-\GGa(\alpha s, \tH-\tz_0)\right) e^{\alpha s(\tH-\tz_0)}. \label{eq:I}
\end{split}
\end{equation}

\begin{table}[bt]
    \centering
    \caption{Coefficients $\PP_{\mp,p}^{(\x)}$ $(p=1,...,4)$, $\PP_{p,q}^{(\x)}$ $p,q=1,2$ in Eq.~\eqref{eq:B} for $\x = \a$, $\b 1$, and $\b 2$. See also Eqs.~\eqref{eq:H} and \eqref{eq:I} for the definitions of $\Iz$, $\IH$, $\dIz$, $\dIH$, $\HH_{j\pm}$, $\HH_{j \up}$ and $\HH_{j \lw}$. }\label{tab:P-tem}
    {
    \begin{tabular}{ll}
    \hline\hline
    \multicolumn{2}{c}{Temperature field}
    \\
    \hline
    & $\x=\a$ (fluid heating)\\
    \cline{2-2}
\multirow{4}{*}{fluid}&  
$\PP_{\mp,1}^{(\a)} =\mp(\HH_{1+}\mp\HH_{1\lw})(\HH_{2+}  \dIH+\HH_{2\up}\IH)e^{\mp s\tz_0}+  (-\HH_{1\lw}\dIz+\HH_{1+}  \Iz)(\HH_{2+}\pm\HH_{2\up})e^{\pm s(\tH-\tz_0)}$\\&
$\PP_{\mp,2}^{(\a)} = - (\HH_{1+}\mp\HH_{1\up})(\HH_{2\lw}\dIH+\HH_{2+}  \IH)e^{\mp s\tz_0}\pm( \HH_{1+}  \dIz-\HH_{1\up}\Iz)(\HH_{2+}\pm\HH_{2\lw})e^{\pm s(\tH-\tz_0)}$\\&
$\PP_{\mp,3}^{(\a)} =\pm(\HH_{1+}\mp\HH_{1\lw})(\HH_{2\lw}\dIH+\HH_{2+}  \IH)e^{\mp s\tz_0}\mp(-\HH_{1\lw}\dIz+\HH_{1+}  \Iz)(\HH_{2+}\pm\HH_{2\lw})e^{\pm s(\tH-\tz_0)}$\\&
$\PP_{\mp,4}^{(\a)} =   (\HH_{1+}\mp\HH_{1\up})(\HH_{2+}  \dIH+\HH_{2\up}\IH)e^{\mp s\tz_0}-  ( \HH_{1+}  \dIz-\HH_{1\up}\Iz)(\HH_{2+}\pm\HH_{2\up})e^{\pm s(\tH-\tz_0)}
$
\\[0.5em]
\multirow{2}{*}{solid 1}&  
$\PP_{1,1}^{(\a)} = [(\HH_{2+}-\HH_{2\lw})(\Iz-\dIz)e^{-s\tH}+(\dIz+\Iz)(\HH_{2+}+\HH_{2\lw})e^{s\tH}-2(\HH_{2\lw}\dIH+\HH_{2+}\IH)]\HH_{1-}$ \\&
$\PP_{1,2}^{(\a)} = [(\HH_{2+}-\HH_{2\up})(\Iz-\dIz)e^{-s\tH}-(\dIz+\Iz)(\HH_{2+}+\HH_{2\up})e^{s\tH}+2(\HH_{2+}\dIH+\HH_{2\up}\IH)]\HH_{1-}$
\\[0.5em]
\multirow{2}{*}{solid 2}&
$\PP_{2,1}^{(\a)} = [-(\HH_{1+}-\HH_{1\lw})(\dIH+\IH)e^{-s\tH}-(\IH-\dIH)(\HH_{1+}+\HH_{1\lw})e^{s\tH}+2(-\HH_{1\lw}\dIz+\HH_{1+}\Iz)]\HH_{2-}$ \\&
$\PP_{2,2}^{(\a)} = [(\HH_{1+}-\HH_{1\up})(\dIH+\IH)e^{-s\tH}-(\IH-\dIH)(\HH_{1+}+\HH_{1\up})e^{s\tH}-2(\HH_{1+}\dIz-\HH_{1\up}\Iz)]\HH_{2-}
$\\ \hline
    & $\x=\b1$ (surface 1 heating)\\
    \cline{2-2} 
\multirow{2}{*}{fluid}&  
$\PP_{1\mp}^{(\b1)} =-  \HH_{1\lw}(\HH_{2+}\pm\HH_{2\up})e^{\pm s(\tH-\tz_0)}$, \quad 
$\PP_{2\mp}^{(\b1)} =\pm 2        (\HH_{2+}\pm\HH_{2\lw})e^{\pm s(\tH-\tz_0)}$\\&
$\PP_{3\mp}^{(\b1)} =\pm\HH_{1\lw}(\HH_{2+}\pm\HH_{2\lw})e^{\pm s(\tH-\tz_0)}$, \quad 
$\PP_{4\mp}^{(\b1)} =-   2        (\HH_{2+}\pm\HH_{2\up})e^{\pm s(\tH-\tz_0)}$ 
\\[0.5em] 
\multirow{2}{*}{solid 1}&  
$\PP_{1,1}^{(\b1)} = (\HH_{2+}-\HH_{2\lw})(\HH_{1\lw}-2) e^{-s\tH}+(\HH_{1\lw}+2)(\HH_{2+}+\HH_{2\lw})e^{s\tH}$ \\&
$\PP_{1,2}^{(\b1)} = (\HH_{2+}-\HH_{2\up})(\HH_{1\lw}-2) e^{-s\tH}-(\HH_{1\lw}+2)(\HH_{2+}+\HH_{2\up})e^{s\tH}$
\\[0.5em] 
\multirow{1}{*}{solid 2}&
$\PP_{2,1}^{(\b1)} = -2\HH_{1\lw}\HH_{2-}$, \quad 
$\PP_{2,2}^{(\b1)} = -4\HH_{2-}$
\\
\hline
    & $\x=\b2$ (surface 2 heating)\\
    \cline{2-2} 
\multirow{2}{*}{fluid}&  
$\PP_{1\mp}^{(\b2)} =\pm 2        (\HH_{1+}\mp\HH_{1\lw})e^{\mp s\tz_0}$, \quad 
$\PP_{2\mp}^{(\b2)} =   \HH_{2\lw}(\HH_{1+}\mp\HH_{1\up})e^{\mp s\tz_0}$\\&
$\PP_{3\mp}^{(\b2)} =\mp\HH_{2\lw}(\HH_{1+}\mp\HH_{1\lw})e^{\mp s\tz_0}$, \quad 
$\PP_{4\mp}^{(\b2)} =-   2        (\HH_{1+}\mp\HH_{1\up})e^{\mp s\tz_0}$ 
\\[0.5em] 
\multirow{1}{*}{solid 1}&  
$\PP_{1,1}^{(\b2)} =  2\HH_{2\lw}\HH_{1-}$, \quad 
$\PP_{1,2}^{(\b2)} = -4\HH_{1-}$
\\[0.5em] 
\multirow{2}{*}{solid 2}&
$\PP_{2,1}^{(\b2)} = -(\HH_{1+}-\HH_{1\lw})(\HH_{2\lw}-2) e^{-s\tH}-(\HH_{2\lw}+2)(\HH_{1+}+\HH_{1\lw})e^{s\tH}$ \\&
$\PP_{2,2}^{(\b2)} =  (\HH_{1+}-\HH_{1\up})(\HH_{2\lw}-2) e^{-s\tH}-(\HH_{2\lw}+2)(\HH_{1+}+\HH_{1\up})e^{s\tH}$
\\
\hline\hline
    \end{tabular}}
\end{table}
\begin{table}[bt]
    \centering
    \caption{Coefficients $\PP_{q\pm,p}^{(\x,\y)}$ $(q=2,3;\,p=1,...,3)$ in Eq.~\eqref{eq:C} for $\x = \a$, $\b 1$, and $\b 2$ and $\y=\c$, $\dd 1$, and $\dd 2$. See also Eqs.~\eqref{eq:J} and \eqref{eq:C} for the definitions of $\Jz$, $\JH$, $\dJz$, $\dJH$, and $\Cxc_{1\pm}$. }\label{tab:P-vel}
    {
    \begin{tabular}{cl}
    \hline\hline
    \multicolumn{2}{c}{Flow field}\\
    \hline 
&$\y=\c$ (thermal convection) \\
    \cline{2-2}
\multirow{6}{*}{$\x=\a$ (fluid heating) }
&$\PP_{2\pm,1}^{(\a,\c)} = \pm\left(\Jz - e^{\pm s\tH}\JH + \dJz +\Cac_{1+}e^{-s\tz_0}+\Cac_{1-}e^{s\tz_0}\right) \sinh s\tH$ \\
&$\PP_{2\pm,2}^{(\a,\c)} = \pm\left(\Jz \cosh s\tH - \JH + \dJH - \Cac_{1+}e^{s(\tH-\tz_0)} - \Cac_{1-} e^{-s(\tH-\tz_0)}\right)$\\
&$\PP_{2\pm,3}^{(\a,\c)} = \JH$ \\ 
&$\PP_{3\pm,1}^{(\a,\c)} = \pm\left(\Jz e^{\mp s\tH} - \JH + \dJH-\Cac_{1+}e^{s(\tH-\tz_0)}-\Cac_{1-}e^{-s(\tH-\tz_0)}\right)\sinh s\tH$ \\
&$\PP_{3\pm,2}^{(\a,\c)} = \pm\left(\Jz - \JH \cosh s\tH + \dJz + \Cac_{1+}e^{-s\tz_0} + \Cac_{1-} e^{s\tz_0}\right)$ \\
&$\PP_{3\pm,3}^{(\a,\c)} = \Jz$,
    \\
    \cline{2-2}
\multirow{6}{*}{$\x=\b 1$, $\b 2$ (surface heating) }
&$\PP_{2\pm,1}^{(\x,\c)} = \pm\left(\Cxc_{1+}e^{-s\tz_0}+\Cxc_{1-}e^{s\tz_0}\right) \sinh s\tH$ \\
&$\PP_{2\pm,2}^{(\x,\c)} = \pm \left(-\Cxc_{1+}e^{s(\tH-\tz_0)} - \Cxc_{1-} e^{-s(\tH-\tz_0)}\right)$\\
&$\PP_{2\pm,3}^{(\x,\c)} = 0$ \\ 
&$\PP_{3\pm,1}^{(\x,\c)} = \pm\left(-\Cxc_{1+}e^{s(\tH-\tz_0)}-\Cxc_{1-}e^{-s(\tH-\tz_0)}\right)\sinh s\tH$\\
&$\PP_{3\pm,2}^{(\x,\c)} = \pm \left(\Cxc_{1+}e^{-s\tz_0} + \Cxc_{1-} e^{s\tz_0}\right)$ \\
&$\PP_{3\pm,3}^{(\x,\c)} = 0$
\\
    \hline 
&$\y=\dd 1$ (surface 1 slip) \\
    \cline{2-2}
\multirow{2}{*}{$\x=\a$, $\b 1$, $\b 2$ }
&$\PP_{2\pm,1}^{(\x,\dd 1)} = \pm s\tH \sinh s\tH$, \quad 
$\PP_{2\pm,2}^{(\x,\dd 1)} = 0$, \quad 
$\PP_{2\pm,3}^{(\x,\dd 1)} = 0$ \\
&$\PP_{3\pm,1}^{(\x,\dd 1)} = 0$, \quad 
$\PP_{3\pm,2}^{(\x,\dd 1)} = \pm s\tH$, \quad 
$\PP_{3\pm,3}^{(\x,\dd 1)} = 0$ 
\\
\hline
&$\y=\dd 2$ (surface 2 slip) \\
    \cline{2-2}
\multirow{2}{*}{$\x=\a$, $\b 1$, $\b 2$ }
&$\PP_{2\pm,1}^{(\x,\dd 2)} = 0$, \quad 
$\PP_{2\pm,2}^{(\x,\dd 2)} = \pm s\tH$, \quad 
$\PP_{2\pm,3}^{(\x,\dd 2)} = 0$ \\
&$\PP_{3\pm,1}^{(\x,\dd 2)} = \pm s\tH \sinh s\tH$, \quad 
$\PP_{3\pm,2}^{(\x,\dd 2)} = 0$, \quad 
$\PP_{3\pm,3}^{(\x,\dd 2)} = 0$ 
\\
\hline\hline
    \end{tabular}}
\end{table}

\subsection{Flow field}
Next, Eq.~\eqref{eq:Stokes-ht} can be solved for $\x=\a$ (fluid heating) and $\x=\b j$ (surface $j$ heating) with $\y=\c$ (thermal convection) and $\y=\dd j$ (surface $j$ slip) as 
\begin{subequations}\label{eq:Stokes-solution}
\begin{align}
&\buzxy\, = \busxy \sum_{\alpha = \pm} \left( \sum_{p=1}^3  \Cxy_{p\alpha}\phi_p + \SSxy(\alpha s, \tz -\tz_0) \right) e^{\alpha s(\tz-\tz_0)}, \\
&\burxy\, = -\frac{1}{s}\busxy \sum_{\alpha = \pm} \left( \sum_{p=1}^3 \Cxy_{p\alpha}\psi_{p\alpha} + \VVxy(\alpha s, \tz -\tz_0)\right) e^{\alpha s(\tz-\tz_0)}, 
\end{align}
\end{subequations}
where
\begin{equation}
\begin{split}
&\phi_1 = \Z (1-\Z), \quad \phi_2 = \Z, \quad \phi_3 = 1-\Z, \quad 
\\
&\psi_{1\alpha} = \frac{1}{\tH}(1-2\Z)+\alpha s \phi_1, \quad 
\psi_{2\alpha} = \frac{1}{\tH}+\alpha s\phi_2, \quad \psi_{3\alpha} =-\frac{1}{\tH}+ \alpha s\phi_3, 
\end{split}
\end{equation}
and the coefficients $\Cxy_{p\pm}$ $(p=1,\,2,\,3)$ and $\busxy$ are the functions of $s$ only; the terms $\SSxy$ and $\VVxy$ are the inhomogeneous solutions expressed by explicit functions of $(s,\tz)$. 
To be more specific, the functions $\busxy$, $\SSxy$, and $\VVxy$ in Eq.~\eqref{eq:Stokes-solution} are defined as, for $\x=\a,\,\b1,\,\b2$,
\begin{equation}
\begin{split}
&
\begin{cases}
\busxc(s) = \frac{1}{2s^2}\tausx(s), \\
\SSac(s,\tz) = \tzR^2\left((\frac{3}{2}-\tzR^2) \GGa(s,\tz) + 
[(\tzR+\frac{s\tz}{2\tzR})^2-\frac{3s\tz}{4 \tzR^2}-1+\frac{3}{4 \tzR^2}] \EEa(s,\tz)\right), \\
\VVac(s,\tz) = s \tzR^2 \left(  [(\tzR+\frac{s\tz}{2\tzR})^2-\frac{s\tz}{2}]\GGa(s,\tz) - [(\tzR+\frac{s\tz}{2\tzR})^2-\frac{s\tz}{4\tzR^2}]\EEa(s,\tz)  \right), \\
\SSbc1(s,\tz) = \SSbc2(s,\tz) = \VVbc1(s,\tz) = \VVbc2(s,\tz) = 0, 
\end{cases}\\
&
\busxd1(s) = s\btx\,(s,0),\quad \SSxd1(s,\tz) = \VVxd1(s,\tz) = 0, 
\\
&
\busxd2(s) = s\btx\,(s,\tH),\quad \SSxd2(s,\tz) = \VVxd2(s,\tz) = 0, 
\end{split}
\end{equation}
where the inhomogeneous solution $\SSac$ and $\VVac$ only appear for the case of fluid heating $(\x=\a)$ and thermal convection $(\x=\c)$. 
The integration coefficients $\Cxy_{p\pm}$ $(p=1,\,2,\,3)$ in Eq.~\eqref{eq:Stokes-solution} depend on $s$ and are symbolically given as, 
for $\x=\a$, $\b 1$, and $\b 2$,
\begin{equation}
\begin{split}
&\Cxc_{1\pm} 
= -\frac{s^2\tH^2}{4}\Bx{\pm}, \quad
\Cxdj_{1\pm} = 0 \quad (j=1,\,2) , 
\\
&\Cxy_{2\pm} 
= \frac{e^{\mp s(\tH-\tz_0)}}{\cosh 2s\tH - 2 s^2\tH^2 -1 }  \sum_{p=1}^3\PP_{2\pm,p}^{(\x,\y)} (s\tH)^{p-1} \quad (\y = \c,\,\dd1,\,\dd2),\\
&\Cxy_{3\pm} 
= \frac{e^{\pm s\tz_0}}{\cosh 2s\tH-2 s^2\tH^2-1} \sum_{p=1}^3\PP_{3\pm,p}^{(\x,\y)} (s\tH)^{p-1} \quad (\y = \c,\,\dd1,\,\dd2), \label{eq:C}
\end{split}
\end{equation}
where $\PP_{2\pm,p}^{(\x,\y)}$ and $\PP_{3\pm,p}^{(\x,\y)}$ $(p=1,\,2,\,3)$ are given in Table~\ref{tab:P-vel} in which $\Jz$, $\JH$, $\dJz$, and $\dJH$ are inhomogeneous terms defined as 
\begin{equation}
\begin{split}
&\Jz  =    \sum_{\alpha=\pm}\SSac(\alpha s,-\tz_0) e^{-\alpha s \tz_0}, \quad 
\JH  =    \sum_{\alpha=\pm}\SSac(\alpha s,\tH-\tz_0) e^{\alpha s(\tH- \tz_0)}, \\
&\dJz = \tH\sum_{\alpha=\pm}\VVac(\alpha s, -\tz_0) e^{-\alpha s\tz_0}, \quad 
\dJH = \tH\sum_{\alpha=\pm}\VVac(\alpha s, \tH-\tz_0)e^{\alpha s(\tH-\tz_0)}. \label{eq:J}
\end{split}
\end{equation}


\section{Detail of the numerical simulation} \label{sec:numerical}
In the following, we describe the numerical method to solve Eqs.~\eqref{eq:temperature-nd} and \eqref{eq:Stokes-nd}, i.e., non-dimensional heat-conduction and Stokes equations with the inhomogeneous terms. 
To avoid redundant notations, we omit the superscript $(\x)$ and $(\x,\y)$. 

\subsection{Overview}
The radial dimension of a computational domain is restricted to $0\leq \tr \leq \tr_{\max}$. When we compare the numerical and semi-analytical solutions, $\tr_{\max}$ should be large enough because the semi-analytical solution is obtained for $\tr_{\max}\to\infty$. 
The $\tz$ variable in the solid 1, the fluid, and the solid 2 are 
discretized at equal intervals using $N_{z1}$, $N_{z}$, and $N_{z2}$ grid points, respectively. In the same manner, the $\tr$ variable is discretized at equal intervals using $N_{r}$ grid points. 
For the five-layer system introduced in Sec.~\ref{sec:nume-anal}, we have additional computational domains inside the thin films, which are discretized in the same manner as above. The number of the grid points for $\tz$ in the thin films 1 and 2 are $N_{m1}$ and $N_{m2}$, respectively.

As we will see in \SI~\ref{sec:stream} below, Eq.~\eqref{eq:Stokes-nd} for the flow field is converted into a form similar to the heat-conduction equation. Therefore, the numerical procedures for Eqs.~\eqref{eq:temperature-nd} and \eqref{eq:Stokes-nd} are overall similar. 
In the fluid domain, a mesh is a rectangule with an area $\Delta \tr \times \Delta \tz$, where $\Delta \tr = \tr_{\max}/N_r$ and $\Delta \tz = \tH/N_z$. 
All the macroscopic quantities (e.g., $\tau$, $u_r$, etc.) are defined on the grid points, and their first and second derivatives are approximated using second-order central difference schemes. Note that we cannot use the central difference schemes in the boundary conditions of the heat-conduction equations Eq.~\eqref{eq:temperature-nd}; we use the second-order forward or backward difference scheme instead.
In the other domains for the solid parts (and for the thin-film parts in the five-layer problem), the same procedure as in the fluid domain is applied. 

We use an iterative procedure based on the Gauss-Seidel method to obtain a steady state. 
The convergence to the steady state is judged by observing whether the maximum value of a residue $|h^{(n+1)}-h^{(n)}|$ ($h=\tau$, $\tau_1$, etc.) over whole grid points is lower than a threshold value of $10^{-16}$, where $n$ is the number of iteration.


\subsection{Stream function-vorticity formulation} \label{sec:stream}
In this subsection, the preliminary procedure, so-called a stream function-vorticity formulation, for solving Eq.~\eqref{eq:Stokes-nd} is introduced. 
We introduce the non-dimensinal stream function $\psi$ so that $\psi$ satisfies $(u_r,u_z) = (\frac{1}{\tr}\rnd{\psi}{\tz},-\frac{1}{\tr}\rnd{\psi}{\tr})$. 
Then, non-dimensional 
vorticity $\omega = \rnd{u_r}{\tz}-\rnd{u_z}{\tr}$ and the stream function $\psi$ satisfy by the following equations:
\begin{align}
\rnd{^2 \omega}{\tr^2} + \frac{1}{\tr}\rnd{\omega}{\tr} - \frac{\omega}{\tr^2} + \rnd{\omega}{\tz^2} -\rnd{\mJ}{\tr} = 0, \quad 
\frac{1}{\tr}\rnd{^2\psi}{\tr^2} - \frac{1}{\tr^2} \rnd{\psi}{\tr} + \frac{1}{\tr}\rnd{^2\psi}{\tz^2} = \omega. \label{eq:streamfunction-vorticity}
\end{align}
As seen below, the boundary conditions of $\omega$ include $\psi$ and thus Eq.~\eqref{eq:streamfunction-vorticity} is a coupled system. 
The boundary conditions for $\psi$ and $\omega$ are summarized as
\begin{subequations}
\begin{align}
&\psi = 0, \quad \omega = \frac{1}{\tr}\frac{2}{(\Delta \tz)^2}(\psi|_{\tz=\Delta \tz}-\tr(\Delta \tz) \mJ_1) \quad (\tz=0,\,0\leq \tr\leq \tr_{\max}), \\
&\psi = 0, \quad \omega = \frac{1}{\tr}\frac{2}{(\Delta \tz)^2}(\psi|_{\tz=\tH-\Delta \tz}+\tr(\Delta \tz) \mJ_2) \quad (\tz=\tH,\,0\leq \tr\leq \tr_{\max}), \\
&\psi = 0, \quad \omega = 0 \quad (0\leq \tz\leq \tH,\,\tr=0), \\
&\psi = 0, \quad \omega = \frac{1}{\tr}\frac{2}{(\Delta \tr)^2}\psi|_{\tr=\tr_{\max}-\Delta \tr} \quad (0\leq \tz\leq \tH,\,\tr=\tr_{\max}), 
\end{align}
\end{subequations}
where the terms including $\mJ_1$ and $\mJ_2$ represent the effect of thermo-osmotic slip flows at the lower and upper interfaces, respectively. 

\subsection{Accuracy checks}
We use the physical parameters in Table~\ref{tab:phys-param} to investigate the accuracy of the numerical simulation. The grid parameters are shown in Table~\ref{tab:num-param}. 
Since the $\tz$ dimension of the fluid domain is $\tH=4$, the grid sizes $\Delta \tr(=\tr_{\max}/N_r)$ and $\Delta \tz(=\tH/N_z)$ in the fluid domain are same in Table~\ref{tab:num-param}.

For case A, we use a grid A1, which is the finest, as a reference grid. 
That is, the relative error between the results obtained by a coarser grid A$j$ ($j=2,3,4$) and that obtained by the grid A1 is investigated. 
We call $\tau_{\mathrm{A} j}^{(\a)}$ the maximum value of $\ta\,$ using the grid A$j$. Then, the indicator of the relative error is defined as $E(\ta\,) = |(\tau_{\mathrm{A} j}^{(\a)}-\tau_{\mathrm{A} 1}^{(\a)})/\tau_{\mathrm{A} 1}^{(\a)}|$. The similar indicators are prepared for the flow speed $|\uac|$, $|\uad1|$, and $|\uad2|$ and are defied as $E(|\uac|)$, $E(|\uad1|)$, and $E(|\uad2|)$, respectively. 
For case B, we use a grid B1 as a reference, and investigate the relative errors of the grid B$j$ ($j=2,3,4$) in the same manner, by introducing the error indicators $E(\tb\,1)$, $E(|\ubc1|)$, $E(|\ubd11|)$, and $E(|\ubd12|)$. 

Figures~\ref{fig:appendix-numerical}(a) and \ref{fig:appendix-numerical}(b) show the $E(X)$ for (a) $X=\ta\,$, $|\uac|$, $|\uad1|$, $|\uad2|$ and (b) $X=\tb\,1$, $|\ubc1|$, $|\ubd11|$, $|\ubd12|$, respectively, as a function of $\Delta \tr(=\Delta \tz)$. It is seen that all the relative-error indicators decrease in proportion to $\Delta \tr^2$. 
This behavior is consistent with our design of the numerical method using the second-order finite difference scheme. Therefore, the method of numerical simulation is validated. 

\begin{figure}
\includegraphics[width=0.7\textwidth]{fig-appendix-numerical\kakuchoushi}
\caption{Relative error $E(X)$ for (a) $X=\ta\,$, $|\uac|$, $|\uad1|$, $|\uad2|$ (case A) and (b) $X=\tb\,1$, $|\ubc1|$, $|\ubd11|$, $|\ubd12|$ (case B). Solid blue lines indicates $\propto (\Delta\tr)^2$. }\label{fig:appendix-numerical}
\end{figure}

\begin{table}[bt]
\centering
\caption{Numerical parameters used in the accuracy checks for the numerical solutions. Note that $\tH(=H/w_0)$, $\tH_1$, and $\tH_2$ are $4$, $8$, and $8$, respectively. }
\label{tab:num-param}
{\tabcolsep = 0.5em
\small
\begin{tabular}{crrrrrrccrrrrrr}
\hline \hline
grid name& $N_r$ & $N_z$ & $N_{z1}$ & $N_{z2}$ & $\tr_{\max}$& $N_{m1}$& &grid name& $N_r$ & $N_z$ & $N_{z1}$ & $N_{z2}$ & $\tr_{\max}$ & $N_{m1}$ \\
A1       & $900$ & $120$ & $120$ & $120$ & $30$ & --- & &B1       & $600$ & $120$ & $120$ & $120$ & $20$ & --- \\
A2       & $600$ &  $80$ &  $80$ &  $80$ & $30$ & --- & &B2       & $400$ &  $80$ &  $80$ &  $80$ & $20$ & --- \\
A3       & $300$ &  $40$ &  $40$ &  $40$ & $30$ & --- & &B3       & $200$ &  $40$ &  $40$ &  $40$ & $20$ & --- \\
A4       & $150$ &  $20$ &  $20$ &  $20$ & $30$ & --- & &B4       & $100$ &  $20$ &  $20$ &  $20$ & $20$ & --- \\
AL       & $240$ & $120$ & $120$ & $120$ &  $8$ & $12$& &BL       & $240$ & $120$ & $120$ & $120$ &  $8$ & $12$\\  
 ALc       & $80$ & $40$ & $40$ & $40$ &  $8$ & $4$& &BLc       & $80$ & $40$ & $40$ & $40$ &  $8$ & $4$\\  \hline
\end{tabular}}
\end{table}






\end{document}